\documentclass[onecolumn]{aastex63}
\newcommand{\be}{\begin{equation}}
\newcommand{\ee}{\end{equation}}
\newcommand{\ba}{\begin{eqnarray}}
\newcommand{\ea}{\end{eqnarray}}
\newcommand{\bnabla}{\mbox{\boldmath$\nabla$}}

\newcommand{\bbeta}{\mbox{\boldmath$\beta$}}
\newcommand{\bxi}{\mbox{\boldmath$\xi$}}
\newcommand{\nn}{\mbox{} \nonumber \\ \mbox{}}

\usepackage{mathtools}

\begin{document}

\title{Radio Emission of Pulsars. I. \\ Slow Tearing of a Quantizing Magnetic Field}

\author{Christopher Thompson}
\affiliation{Canadian Institute for Theoretical Astrophysics, 60 St. George Street, Toronto, ON M5S 3H8, Canada}  

%\date{\today}
\shortauthors{Thompson}
\shorttitle{Radio Emission of Pulsars. I. Slow Tearing of a Quantizing Magnetic Field}
%\submitjournal{The Astrophysical Journal}
\accepted{29 January 2022}

\begin{abstract}
  The pulsed radio emission of rotating neutron stars is connected to slow tearing instabilities
  feeding off an inhomogeneous twist profile within the open circuit. This paper considers
  the stability of a weakly sheared, quantizing magnetic field in which
  the current is supported by a relativistic particle flow.   The electromagnetic field is
  almost perfectly force-free, and particles are confined to the lowest Landau state,
  experiencing no appreciable curvature drift.   
  In a charge-neutral plasma, %(equal densities of electrons and positrons)
  we  find multiple branches of slowly growing tearing modes, relativistic analogs of the double tearing mode, with
  peak growth rate $s \gtrsim 4\pi \widetilde k_y J_z/B_z$.  Here, $B_z$ is the strong (nearly potential) guide magnetic field,
  $J_z$ the field-aligned current density, and $\widetilde k_y$ is the mode wavenumber normalized by the current gradient
  scale.  These modes are overstable when the plasma carries net charge,
  with real frequency $\omega \sim s\cdot |n_0^+ - n_0^-|/(n_0^+ + n_0^-)$ proportional
  to the imbalance in the densities of positive and negative charges.   An isolated current sheet
  thinner than the skin depth supports localized tearing modes with growth rate scaling as
  (sheet thickness/skin depth)$^{-1/2}$.   In a pulsar, the peak growth rate is comparable to the angular
  frequency of rotation, $s \gtrsim 2\widetilde k_y  \Omega$, slow compared with the longitudinal oscillations
  of particles and fields in a polar gap.   The tearing modes experience azimuthal drift reminiscent
  of sub-pulse drift and are a promising driver of pulse-to-pulse flux variations.  
  A companion paper demonstrates a Cerenkov-like instability of current-carrying Alfv\'en waves in thin
  current sheets with relativistic particle flow, and proposes coherent curvature emission by these waves
  as a source of pulsar radio emission.
  \end{abstract}


\section{Introduction}

An outstanding question about stellar magnetic fields centers on the role that internal resistive instabilities play
in redistributing magnetic twist and the associated electric current.  This phenomenon has
long been encountered in magnetic plasma confinement devices (tokamaks), where an elaborate theory of these
instabilities has been developed \citep{white13}.  High-order resistive instabilities involving multiple magnetic tearing
surfaces are known to play a central role in establishing the large-scale twist profile within the plasma torus.
This process can be viewed, in some respects, as a relaxation to a lower-energy state while conserving magnetic helicity \citep{taylor86}.

As regards stellar magnetospheres, most studies of magnetic tearing have focused on the effect of divergences
in the magnetic shear that is imposed at the surface of the star, or on collisions between discrete magnetic
structures \citep{parker72,priest00,nalewajko16}.
A few investigations have been made of the microscopic underpinnings
of turbulent relaxation in stellar magnetospheres (e.g. \citealt{browning86,boozer20}), but none directed toward the relativistic
magnetosphere of a neutron star.

The case of a neutron star presents unique challenges and interesting opportunities.  The main thesis of this paper
is that the twist profile imposed on a pulsar magnetosphere by the global charge flow is susceptible to internal tearing
instabilities.  The open magnetic flux bundle is very narrow near the star; high-order instabilities are
less sensitive to the global connectivity of the magnetic field,
and in principle will arise even in zones that do not sustain electron-positron pair creation.  We find that internal
tearing in the inner pulsar magnetosphere is slower
than the plasmoid instability seen near the magnetic separatrix in numerical simulations of the outer magnetosphere
and wind zone \citep{cerutti17,philippov19}.
The characteristic scale is small -- around the plasma skin depth of $\sim 0.1-100$ cm -- and the growth time
comparable to the rotation period of the star.

A companion paper \citep{t21} identifies a secondary, Alfv\'enic instability of
small-scale current sheets, which shows promise in explaining the charge clumping that is needed to sustain coherent
curvature emission in the radio band.

\subsection{Tearing in a Quantizing Magnetic Field}

This paper has a few interconnected goals.  The first is to extend the theory of magnetic reconnection
in a collisonless, relativistic plasma to the quantizing regime, where a strong guide magnetic field
is present, the magnetization may approach $\sim 10^{20}$,
and the effects of particle gyromotion and curvature drift are essentially negligible.  We adopt
a kinetic approach, as this problem cannot reliably be approached by a relativistic form of resistive
magnetohydrodynamics.  Previous kinetic investigations of non-relativistic \citep{galeev76,drake_lee77,basu81,ottaviani93}
and relativistic \citep{zelenyi79,zenitani08,sironi11,cerutti13,sironi14,hoshino21} reconnection have focused on plasma configurations
where curvature drift is not negligible, either because particles gain relatively high energies, or because
the guide magnetic field is not very strong.

In the relativistic regime with a strong
guide field, we find that a localized tearing mode can be excited
at an isolated current sheet if the sheet thickness is comparable to or smaller than the skin depth.
In addition, we allow the plasma to have a finite charge density,
finding that the tearing instability becomes an overstability:  magnetic islands experience a general
drift along the tearing surface.

A second goal is to build theoretical tools that will be useful in studying
the spontaneous relaxation of a smoothly twisted magnetic field
in a stellar magnetosphere.  This process is, to a limited extent, constrained by the tying of magnetic field
lines at the stellar surface;  the endpoint is generally not a simple linear
force-free magnetic field, with current density and magnetic flux density connected by a spatially uniform constant
(e.g. \citealt{russell15}).
A smoothly twisted field is, nonetheless, generally susceptible to some degree of relaxation.  In a tokamak,
where the relaxation process is
less well described by Taylor theory than in a reversed-field pinch, configurations with magnetic twist peaking
off the toroidal axis will quickly relax to a more centrally peaked configuration.  Internal tearing modes feeding
off multiple interacting current sheets appear to play a significant role.  They are a major focus of our study.

  The plasma in the magnetosphere of a neutron star is so dilute that electron inertia can play a stronger
  role than Coulomb collisions in mediating the breakdown of magnetic flux freezing.  We also focus on the
  case where ions are absent, corresponding to a negative corotation charge in the magnetosphere.
  The ordering of the electron gyro-scale and Debye scale is reversed in the relativistic plasma,
  as compared with the more familiar non-relativistic case.  We may therefore contrast our approach with 
  \cite{galeev76}, who in their treatment of non-relativistic  reconnection considered tearing modes
  that are strongly influence by electron gyromotion, even while the magnetic gradients are restricted to
  scales much exceeding
  the Debye length.  In a more complete treatment of non-relativistic reconnection
  \citep{drake_lee77, basu81}, both the electron gyroscale and
  Debye scales are resolved, but the tearing eigenfunction incorporates an outer ideal zone where ions
  provide a neutralizing charge.  In contrast with the relativistic case,
  the electrostatic potential contributes minimally to the parallel electric field.
  \cite{ottaviani93} give a simplified treatment of interacting tearing surfaces that resolves the
  Debye scale and neglects collisions, but also incorporates heavy, neutralizing ions.  

  In the case of a relativistic and dilute plasma, where collisions are slow and ions are absent, we find that
  the outer tearing mode never fully reaches the ideal regime.  Growing modes are obtained in the presence
  of multiple interacting (periodic) tearing surfaces.  Similarly to the collisionless, non-relativistic case
  (e.g. \citealt{ottaviani93}), we find that when the separation between tearing surfaces much exceeds the
  Debye scale, the approximation of a uniform magnetic flux perturbation breaks down near
  each tearing surface;  in a related effect, the mode displacement is strongly concentrated in the bulk.
  However, we also find a tower of
  overtone modes with finite wavenumber transverse to the current sheet and
  growth rates nearly as large as the fundamental mode.  These long range oscillations in the
  perturbed magnetic field are ascribable to the absence of heavy ions and breakdown of screening.

%\cite{galeev76} considered perturbations of a much longer wavelength
%than the Debye length, neglecting space charge effects.  \cite{ottaviani93} resolved the Debye length, but included
%the neutralizing effects of heavy ions and also neglected space charge effects.

\subsection{Pulsars and Magnetars}

Our third goal is to investigate the role that the spontaneous relaxation of twist may play in sourcing
radiative emissions from a neutron star.  Twist is injected into the magnetosphere of a radio pulsar
by an escaping flow of corotation charge \citep{gj69,spitkovsky06};
in a magnetar, by an internal magnetoelastic instability \citep{tlk02,chen17}.

The radio emission of rotating neutron stars exhibits (i) very short-timescale variations
(microstructure), implying strong angular variations within a single pulse \citep{graham-smith03}; and (ii) much slower 
pulse-to-pulse variations in the envelope of a radio pulse, including a secular drift in the phase
distinct sub-pulses (e.g. \citealt{deshpande01}).
The tearing mode analysis presented here makes some suggestive connections with these well-known properties of pulsars.
First, tearing is found to be concentrated on small lengthscales, comparable to the skin
depth of $\sim 0.1-100$ cm  in the open pulsar circuit.  Second, the growth of the tearing instabilities
is constrained by the strong background dipolar magnetic field, being limited to a rate comparable to the
angular frequency of rotation.  In addition, high-order tearing modes in a nearly azimuthally symmetric plasma
can form a discrete ring of current `knots' \citep{bierwage05,wang07}.  
The tearing process turns out to be an overstability in the presence of a finite charge density,
as is expected in the pulsar magnetosphere.   Angular drift is therefore an intrinsic property of such
structures.

The main implication of this work for pulsars is that the process of radio emission is generally
top-down, involving internal tearing of a large-scale magnetic field.   It has long been
suspected that the emission of Crab-like pulsars arises from the separatrix between open and closed magnetic
field lines, or the equatorial current sheet in the wind (e.g. \citealt{gruzinov07,cerutti17}).
Internal tearing modes are slowly growing in the open circuit, but
%The resistive instabilities occuring in the open circuit are more slowly growing, but
still potentially play a potent role in adjusting the current profile over a single rotation.

The magnetosphere of a magnetar appears to sustain a relatively stronger current than is encountered in a
rotation-powered pulsar.
The instabilities found here also provide a starting point for investigating the
the redistribution of currents within the magnetosphere, which is mediated by a combination of internal tearing
\citep{thompson08} and ohmic dissipation \citep{beloborodov09}, the relative importance of which is not understood.

\subsection{Plan of the Paper}

The plan of this paper is as follows.  Section \ref{s:quant} sets the stage by reviewing the plasma properties
expected in the magnetosphere of a radio pulsar or magnetar, and some basics about internal tearing in a magnetized
plasma.  Next, in Section \ref{s:ff}, we clarify the connection between charge conservation along relativistic magnetic flux tubes and
vortical deformations of a nearly force-free and quantizing magnetic field.   Our kinetic approach to linear perturbations
is described in Section \ref{s:kinetic}.  Growth rates are derived in Section \ref{s:neutral} for linear tearing modes
in a neutral plasma supporting a periodic flux component with a wavelength somewhat larger than the skin depth.   This
result is generalized in Section \ref{s:single} to a single current sheet.  A shift from instability to overstability in
a plasma with net charge density is demonstrated in Section \ref{s:nonneutral}. The effect of line tying on
the tearing mode and the plastic deformation of the upper neutron star crust are discussed in Section
\ref{s:linetying}.   A summary of our results is presented
in Section \ref{s:summary}, along with some implications for radio pulsars and magnetars.

Throughout this paper, we adopt the shorthand $X = X_n\times 10^n$ to describe the normalization of quantity $X$
in c.g.s. units.

\section{Collisionless Plasma in a Quantizing Magnetic Field}\label{s:quant}

In this paper, we investigate the collective dynamics of charged particles and a 
quantizing magnetic field ($B \gtrsim 10^8$ G in radio pulsars and $\gtrsim 10^{14}$ G in magnetars).
The particles relax rapidly to the lowest Landau state and, since ${\bf B}$
is deformed only slowly, are guided along this field.  In other words, one may locally define a frame where
the electric field ${\bf E} \parallel {\bf B}$ and $E^2 \ll B^2$,
although globally this is not possible.  The particle flow is
driven by a gentle transverse shearing of the magnetic field.  The self-inductance of the system is enormous:
the non-potential component of the magnetic field may carry several orders of magnitude more energy than the charges.
The particle distribution function is locally one-dimensional and highly nonthermal.

The ratio of plasma frequency
to gyrofrequency is extremely small.  In a frame where the plasma has vanishing net momentum,
\be
   {\omega_p\over\omega_c} = {(\langle\gamma^{-3}\rangle 4\pi n q^2 / m)^{1/2}\over |q|B/\gamma m c}
   \sim \sigma^{-1/2},
\ee
where $\sigma = B^2/4\pi (n\gamma mc^2)$ is the magnetization, $\gamma$ is a characteristic Lorentz factor,
and $q$, $m$, $n$ are the particle charge, mass and number density.  We will therefore neglect curvature
drift and other effects of gyromotion.

We consider two cases: (i) a charge-balanced plasma (charges
$\pm q$ are present in equal numbers) and (ii) a plasma composed of charges of a single sign, with net
charge density $\pm |q|n$ (as appropriate to some parts of the pulsar magnetosphere where $e^\pm$ pairs
may be relatively rare).

Our analysis of tearing instability focuses on the simplest case, where $m$ and $|q|$ are the same for
all plasma particles.  After deriving the basic kinetic equations, we first consider the simplest case
of a charge-balanced plasma (Sections \ref{s:neutral} and \ref{s:single}).
Next, deviations from charge neutrality are treated
as a perturbation to the charge-neutral case (Section \ref{s:nonneutral}).

\subsection{Pulsars}

Our approach is motivated by the application to radio pulsars and magnetars.  When a neutron star rotates
with frequency $\Omega$, its corotating
magnetosphere supports a net charge density $\rho_c \sim \mp\Omega B/2\pi c$.
Allowing for ${\cal M}_\pm$
pairs per corotation charge (and for the possibility that the seed charges are protons rather than electrons) gives
\ba
\sigma &=& \left(\gamma m + 2{\cal M}_\pm \gamma_\pm m_e\right)^{-1}\,{|e|B\over 2\Omega c}\nn
&=& {1.4\times 10^{17}\over \gamma m/m_e + 2{\cal M}_\pm \gamma_\pm} B_{12} P_{-1},
\ea
where $P = 2\pi/\Omega$ is the spin period and $-e$ and $m_e$ are the electron charge and mass.
The plasma skin depth is some 7 to 9 orders of magnitude
larger than the gyroscale, and is given by
\be
k_p^{-1} = {c\over \omega_p} = {c\over (2\omega_{c,e} \Omega)^{1/2} g}
= {0.64\over g}\,B_{12}^{-1/2} P_{-1}^{1/2} \quad {\rm cm},
\ee
where $g(m, \gamma, \gamma_\pm, {\cal M}_\pm) =
[(m_e/m)\langle\gamma^{-3}\rangle + 2{\cal M}_\pm\,\langle\gamma_\pm^{-3}\rangle]^{1/2}$ and $\langle...\rangle$
denotes an average over the one-dimensional momentum distribution.
The parts of the polar magnetic flux bundle where $k_p^{-1} \sim 0.1-10^2$ cm are of greatest interest
for radio emission.

Models of nebular synchrotron radio emission around young, rapidly spinning pulsars provide 
constraints on the pair multiplicity ${\cal M}_\pm$, in at least some parts of the pulsar circuit.
In the case of the Vela pulsar, ${\cal M}_\pm \sim 10^5$ is inferred \citep{dejager07,bucciantini11}.
Theoretical models of pair creation in the polar cap of a neutron star depend on the self-consistent
solution for ${\bf E}\cdot{\bf B}$, and imply a range of multiplicities \citep{hibschman01,th19}.
The outcome of these models is influenced by secondary effects such as relativistic frame dragging
\citep{mt92,philippov15}, and will also be influenced by changes in the local current density driven
by the instability described here.

 The role of Coulomb collisions in magnetic tearing in the pulsar circuit depends in interesting
  ways on ${\cal M}_\pm$, as well as on the strength of the magnetic field relative to the QED field
  $B_{\rm QED} = m_e^2c^3/e\hbar  = 4.4\times 10^{13}$ G.  Collisions that are trans-relativistic
  in the center-of-momentum frame have a total cross section comparable to the unmagnetized value
  when $B \ll B_{\rm QED}$.  Comparing the collision rate $s_{\rm col}$ with a mode growth rate $s \sim
  4\pi|J|/B$,
  one finds $s_{\rm col}/s \sim \alpha_{\rm em}(B/B_{\rm QED}){\cal M}_\pm$ when $J \sim \rho_cc$
  (and the corotation charge is supplied by electrons).    Thus, collisions will have
  a negligible effect in the parts of the pulsar circuit where the $e^\pm$ multiplicity is modest, and
  especially where the particle kinetic energy remain below threshold for triggering a $e^\pm$ pair
  cascade.  In Paper II, we show that matching the drift rate of internal tearing
  modes with the angular drift rate of radio sub-pulses suggests low or vanishing ${\cal M}_\pm$
  within the drifting magnetic structures.

  A new channel for $e^\pm$ backscattering opens up when $B \gtrsim 5\,B_{\rm QED}$:  a single
  gamma ray produced by $e^\pm$ annihilation will quickly re-convert back to a pair, $e^+ + e^- \rightarrow
  \gamma \rightarrow e^+ + e^-$ \citep{tk20}.  This is mediated by the resonant quantum-electrodynamic
  $s$ channel, with
  a cross section $\sim 10^2$ times larger than for the Coulomb-like $t$ channel.    This suggests
  that collisions will have a stronger effect on internal tearing in the magnetosphere of a magnetar
  than in ordinary radio pulsars.

\subsection{Magnetars}

So far, two complementary models have been developed for the plasma state surrounding a magnetar.  It is possible
that a counterstreaming flow of electrons and positrons is sustained in a double-layer structure
by the resonant scattering of blackbody keV-energy photons flowing from the heated stellar surface \citep{bt07}.
In this case, the plasma is nearly collisionless and the $e^\pm$ flow is
close to what is required to sustain a large-scale (e.g. hemispheric) magnetic twist,
$n_\pm \simeq |\bnabla\times{\bf B}|/4\pi e$.  Hence, the magnetization $\sigma \sim eB\ell_B/\gamma m_ec^2
\sim 10^{16} B_{15}\ell_{B,7}/\gamma_3$, where $\ell_B = B/|\bnabla\times{\bf B}|$ is the magnetic twist length
and $\gamma \sim 0.03 \hbar eB/m_e c kT_{\rm bb} \sim 10^3\,B_{15}(kT_{\rm bb}/{\rm keV})^{-1}$ is
the Lorentz factor of the resonantly interacting $e^\pm$.

A recent re-examination of QED processes in ultrastrong magnetic fields has revealed the possibility
of a collisional plasma state with a high resistivity, unlike the most common picture
of a pulsar magnetosphere \citep{tk20}.  This collisional state would be sustained
by frequent pair annihilation and reconversion, $e^+ + e^- \rightarrow \gamma \rightarrow
e^+ + e^-$.  The required current density could be sourced by yielding of the magnetar
crurst in compact fault-like zones (as suggested by ab initio calculations of magnetic
diffusion: \citealt{gourg16,tyo17}).  The state of the $e^\pm$ plasma is
trans-relativistic, quasi-thermal and stable, with density $n_\pm \sim 15\,|\bnabla\times{\bf B}|/4\pi e$
and magnetization about $10^2$ times higher than
in the relativisitic double-layer model.   One also obtains a direct explanation
for the presence of a hard X-ray component of the magnetar spectrum:  
the annihilation of trans-relativisitic $e^+-e^-$ pairs in magnetic fields $B \gtrsim 10^{14}$ G produces a broad,
bremsstrahlung-like spectrum of X-rays \citep{tk20}, similar to that observed.
The higher density in this state  also provides
a promising context for collective plasma emission in the IR-optical band, which is observed from quiescent
magnetars at a rate far exceeding the expected surface blackbody flux \citep{kb17}.

\begin{figure}
  \epsscale{1}
  \vskip 0.2in
  \plottwo{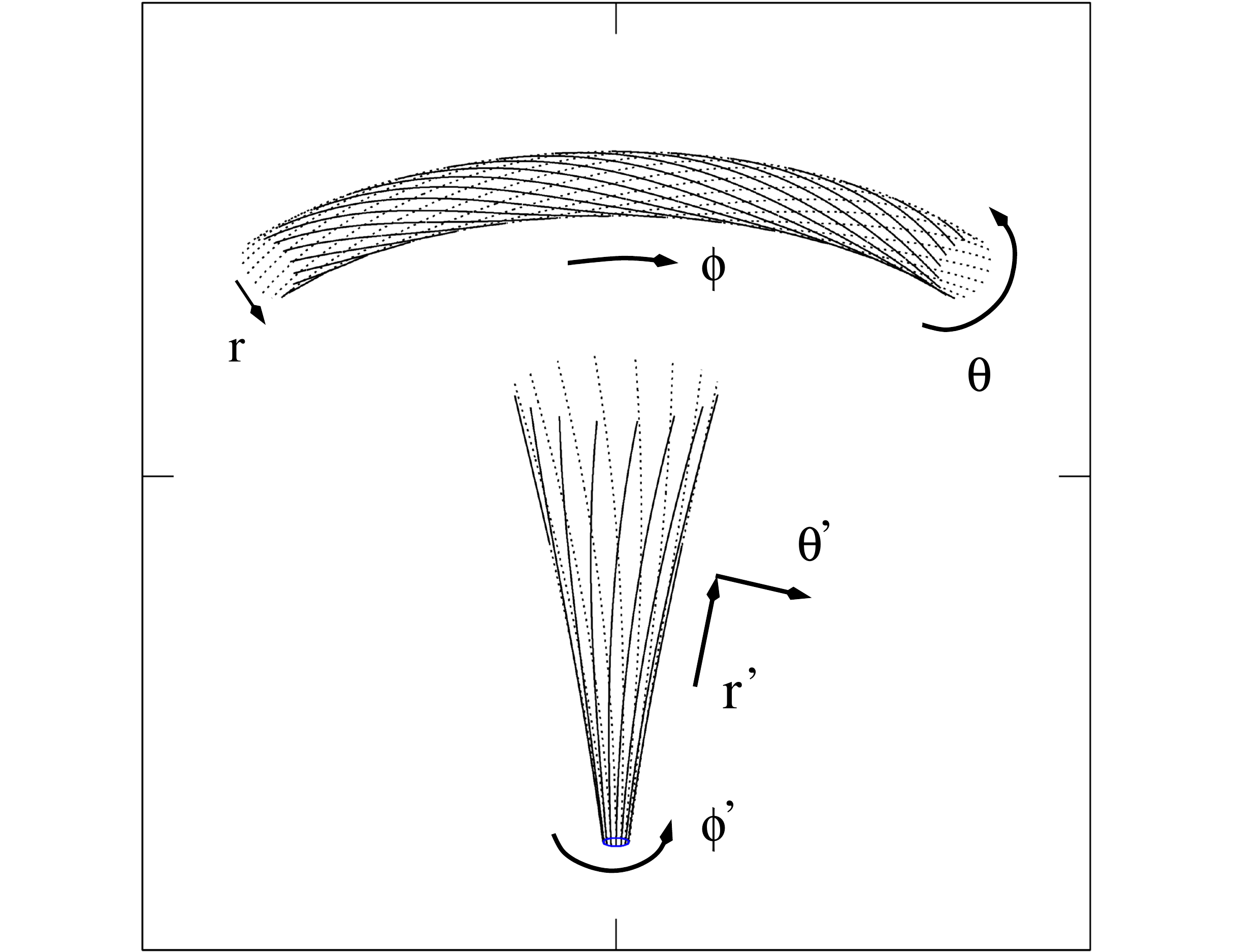}{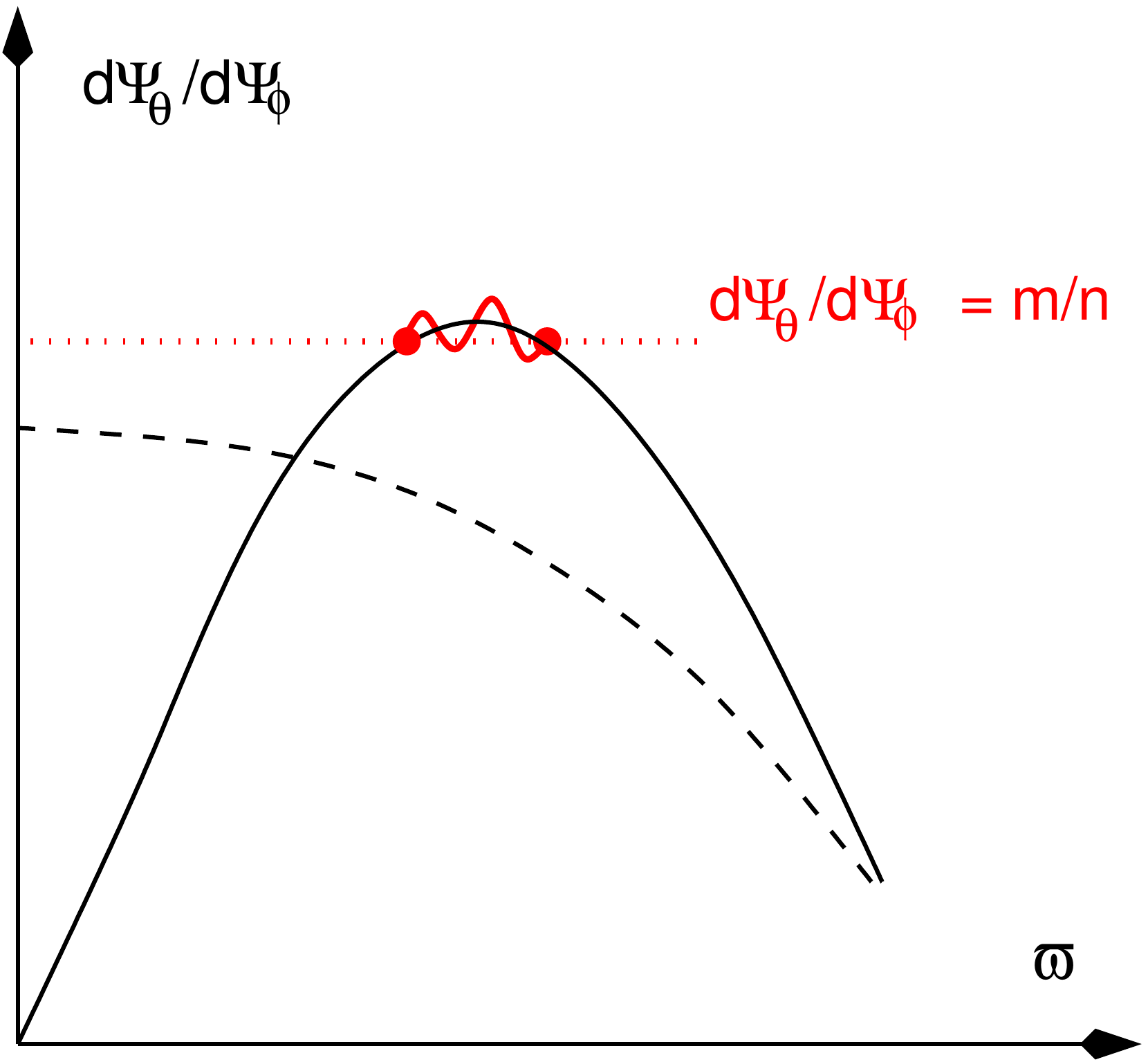}
  \vskip .2in
\caption{{\it Left panel:} Magnetic flux surface in a current-carrying plasma.
  Top:  toroidal geometry (tokamak).   Bottom:  polar magnetic flux bundle of a radio pulsar.
  {{\it Right panel:}  Evolution of the twist profile in toroidal geometry.   High-order
    tearing modes associated with overlapping rational surfaces are triggered when the twist peaks off the main toroidal axis.
    The analysis in this paper focused on the stability of high-wavenumber current gradients that are superimposed
    on the mean helical magnetic field. We consider}
  %  The analysis  of tearing instability in this paper focuses on
  a local cartesian patch, with guide magnetic field
  oriented in the $\hat z$ direction (corresponding approximately to direction $\hat\phi$ in the torus
  and $\hat r$ in the open circuit of a radio pulsar).  %We investigate tearing modes that feed off 
  %high-wavenumber magnetic flux variations superimposed on the large-scale magnetic twist.
  In a magnetic confinement device, these modes drive the redistribution of twist across the small axis of
  the torus;  in a radio pulsar, we propose that they mediate a release of non-potential magnetic field energy
  and are the underlying instability driving coherent radio emission.\label{fig:torus}.  The linear stability analysis
    presented here is combined with a conjecture that the redistribution of twist across magnetic flux surfaces is
    driven by small-scale current variations, extending down to scales comparable to the skin depth.}
\vskip .3in
\end{figure}

\subsection{Internal Tearing in Cylindrical and Toroidal Geometry}\label{s:toroidal}

In this section, we first review some basic aspects of internal tearing in a magnetized plasma with toroidal
or cylindrical symmetry, which may be useful for an astrophysical audience.
Then we make contact with the non-potential magnetic field configuration encountered
in the polar regions of a neutron star magnetosphere.  Our focus in this paper is on the interaction of multiple,
closely spaced tearing surfaces, which can be studied using a (locally) cartesian toy model.   The analysis in
Sections \ref{s:ff}-\ref{s:nonneutral} is based on such a cartesian model.

A simple point of notational confusion may arise here (Figure \ref{fig:torus}).  Distinct coordinate systems are adopted
in studies of pulsar electrodynamics, where the active magnetic field lines are concentrated in a narrow
bundle near the magnetic dipole axis, and in the literature dealing with the breakdown of confinement
in toroidal plasmas (tokamaks).  In the first case, standard spherical polar coordinates are frequently adopted.
In the latter case, the toroidal magnetic field $B_\phi$ winds around the large radius of the torus, and
the poloidal field $B_\theta$ around the small radius.

In toroidal geometry, the unperturbed magnetic field can be written
\be
   {\bf B}_0 =  B_{\phi0}(r,\theta)\hat\phi + B_{\theta0}(r,\theta)\hat{\theta} = 
   \bnabla\Psi_\phi\times\bnabla\theta - \bnabla\Psi_\theta\times\bnabla \phi,
\ee
where $r$ is a locally cylindrical radial coordinate (Figure \ref{fig:torus}).
This expression manifestly satisfies $\bnabla\cdot{\bf B}_0 = 0$, and is useful for analyzing
current-driven instabilities.
The flux functions $\Psi_\phi$, $\Psi_\theta$ are both constant along the magnetic field:
${\bf B}_0\cdot \bnabla\Psi_\phi = {\bf B}_0\cdot \bnabla\Psi_\theta = 0$.  In the case of a slender torus,
the poloidal field $B_{\theta0}$ does not depend significantly on the large radius $R_c$ of the torus, which
enters this equation via $\bnabla\phi = \hat\phi/R_c$.   It is therefore sometimes useful to write fields and
currents in terms of the linear poloidal flux, $\psi_\theta = \Psi_\theta/R_c$.

The magnetic twist is commonly described in terms of a `safety' factor:
\be
   {1\over q} = {d\Psi_\theta\over d\Psi_\phi}.
\ee
A toroidally symmetric equilibrium corresponds to $q = q(r)$.  Large $q$ (strong `safety') corresponds to slow rotations of the
magnetic field lines in the $\theta$-direction.  This regime of weak magnetic twist has a direct analog
in the open pulsar circuit.

Enhanced tearing is encountered in configurations in which $q(r)$ has extrema at $r>0$, and can be triggered
even when $1/q$ is small (the twist is weak).
Important for us is the ability of high-wavenumber oscillations in $q(r)$ to induce tearing, even
when superposed on a smooth $q$ profile that dominates the current density.  This is typically demonstrated by
identifying the mean helical flux surface $\bar\Psi_\theta$,
\be
\Psi_\theta = \bar\Psi_\theta + \Delta\Psi_\theta,
\ee
and then considering a linear mode with wavevector ${\bf k}$ perpendicular to this surface.
The remaining flux component $\Delta\Psi_\theta$ sources a more strongly variable field,
$\Delta B_\theta =  R_c^{-1} (d\Delta\Psi_\theta/dr) \hat\theta$, which may exhibit
multiple null surfaces $\Delta B_\theta = 0$.

An essential feature of the tearing process is that the perturbation equations directly involve the background
current density only through its gradient \citep{biskamp05}.   This property holds also in the nearly force-free regime,
as will become clear in Section \ref{s:kinetic}.
As long as the gradient scale of $\bar\Psi_\theta$ is much longer than that of $\Delta\Psi_\theta$,
the mean current does not significantly modify the growth of the high-wavenumber tearing mode.   This is seen directly
in our kinetic approach, which follows the perturbed distribution function of the current-carrying charges, whose
overall drift speed is fixed by the mean magnetic twist.

\subsubsection{High-wavenumber Tearing in Cylindrical Symmetry}\label{s:cyl}

With a goal of investigating internal tearing in the pulsar emission zone, consider now a very slender torus
and focus on a small segment of length $\Delta z = R_c\Delta\phi \ll R_c$.
A locally cylindrical coordinate system $(r, \theta, z)$ is adopted.   The torus may experience
long-wavelength deformations that change $R_c$ and break toroidal symmetry.   This complicates
low-wavenumber tearing modes but can have a weaker impact on high-wavenumber modes that feed off small-scale
gradients in the magnetic field.

For now, we maintain toroidal symmetry
in the background plasma state (invariance under rotations in $\phi$) and look for linear modes depending on the angular variable
\be
m\theta - n\phi \;=\; m\theta - \left({n\over R_c}\right)R_c\phi \;=\; m\theta - k_z z,
\ee
with wavevector
\be
   {\bf k} = \bnabla(m\theta - n\phi) = {m\over r}\hat\theta - k_z \hat z.
\ee
This is chosen to be orthogonal to the mean helical magnetic field, ${\bf k}\cdot\bar{\bf B}_0 = 0$.
The unperturbed magnetic field decomposes as 
\be
{\bf B}_0 = \bar{\bf B}_0 + \Delta{\bf B}_0 = \bar B_{z0}\left(\hat z + {rk_z\over m}\hat\theta\right) -
     \bnabla(\Delta\psi_\theta)\times\bnabla z,
\ee
where $B_{z0} = B_{\phi 0}$ varies weakly across the flux bundle in the force-free regime.
We will take $m$ and $k_z$ both to be large, given that $\Delta{\bf B}_0$ varies on a small scale. 

The flux variable decomposes similarly as
\be
   \psi_\theta \;=\; \bar\psi_\theta + \Delta\psi_\theta \;=\; {r^2\over 2m}k_z\bar B_{z0} + \Delta\psi_\theta.
\ee
The unperturbed current profile is then
\be
   {4\pi\over c}{\bf J}_0 = \left[{2k_z\over m}B_{z0} + \nabla^2_\perp(\Delta\psi_\theta)\right]\hat z
      - {\partial B_{z0}\over\partial r}\hat\theta,
\ee
where $\nabla_\perp^2$ is the Laplacian in the $(r,\theta)$ plane.
The first term in $J_z$, which is sourced by the mean helical field, varies smoothly and
largely factors out of the tearing mode analysis.  The $\theta$-component of ${\bf J}_0$ is needed
to maintain force-free equilibrium (see Section \ref{s:ff}).

\subsubsection{Neutron Star Magnetosphere}

Now consider the polar region of a neutron star magnetosphere (bottom panel of Figure \ref{fig:torus}), through which a current is driven either
by the corotation charge flow \citep{gj69}, or by a subsurface instability, as in the case of magnetars \citep{tlk02}.  A change of
coordinates and notation is now required.  The guide magnetic field is identified not with the toroidal
component (as in the torus geometry), but with a dipole field centered on the star.  We adopt spherical polar coordinates
$(r',\theta',\phi')$.   Near the magnetic dipole axis, this poloidal field is
approximately radial, with a small flaring
$B_{\theta'} \simeq {1\over 2} \theta' B_{r'}$.  Within the corotating magnetosphere, a flux surface anchored
at polar angle $\theta_s'$ on the star will reach a maximum radius $r'_{\rm max} = r'_s/\sin^2\theta_s'$
(here $r_s'$ is the stellar radius);  its curvature weakens with increasing radius, $R_c(r') \sim (r' r'_{\rm max})^{1/2}$.
Magnetic field lines extending beyond the corotating magnetosphere are confined to an approximately
cylindrical bundle of opening angle $\theta'_{\rm open}(r') \simeq (r'\Omega/c)^{1/2}$, with a radius of curvature
$R_c \sim (r'c/\Omega)^{1/2}$.  The curvature of the open field lines experiences mild oscillations during a single
rotation of the star.

The weak magnetic field driving reconnection is now the (approximately) toroidal component.  Consider, first,
the corotating part of the magnetosphere, within which a flux surface may be twisted through an angle
$\Delta\phi'$ between the north and south polar regions.   This twist angle generally depends on the footpoint colatitude
$\theta_s'$.   The toroidal magnetic field $B_{\phi'} \simeq {1\over 2}\Delta\phi' \theta'^{\,2}B_{\theta'}
\simeq {1\over 4}\Delta\phi' \theta'^{\,3} B_{r'}$ is supported by the current density
$J_{r'} = (c\theta'^{\,2}B_{r'}/4\pi r')\Delta\phi'$.

On the open-field bundle, the toroidal field is sustained by the corotation charge flow $J_{r'} \simeq \rho_{\rm co}c$,
where $\rho_{\rm co} \simeq -\Omega B_{r'}/2\pi c$.   This outward flow is compensated by a return current flowing
through an annular sheath surrounding the polar cap.
In contrast with the current supporting a closed-field magnetic twist, the rotation-driven current $J_{r'}$ has
the same sign in each hemisphere.

Consider the twist profile as a function of transverse radius $\varpi' = \theta' r'$ in the simplest case of an axisymmetric
current flow.  Writing $J_{r'}(\varpi') = J_{r'}(0)[1 - f(\varpi')]$, one finds
\be
   {1\over qR_c} = {|B_{\phi'}|\over \varpi' B_{r'}}  = {4\pi\over B_{r'}c(\varpi')^2}\left|\int_0^{\varpi'} dw' w'
     J_{r'}(w')\right| = {|\Omega|\over c}\left[1 - {1\over(\varpi')^2}\int_0^{\varpi'}dw'w'f(w')\right].
\ee
In the simplest case where a central core of the open flux bundle maintains a flat current density, one sees that the
profile of $qR_c$ is also flat.   The tearing instability considered here therefore feeds off gradients in $f(\varpi')$.
Otherwise, the profile of $qR_c$ peaks near $\varpi' = 0$.

Three-dimensional, force-free models
of the pulsar magnetosphere reveal strong departures from uniform current density near the center of the open flux
bundle, as well as significant departures from axial symmetry \citep{timokhin13,gralla17}.
The azimuthally averaged twist profile generally shows an off-axis maximum, which in a tokamak is known to trigger
a rapid tearing instability that is driven by the interaction between rational surfaces \citep{white13}.

In the case of magnetars, global models of yielding in the solid crust suggest that twist is injected
in localized, fault-like zones \citep{gourg16,tyo17}.  The inhomogeneities in the excited twist are then even
strong than those that are imposed by spindown in the open circuit.  Tearing modes similar to
those investigated here are implicated in the broader spreading of the current through
the closed magnetosphere \citep{thompson08}.

\vfil\eject
\section{Nearly Force Free Dynamics}\label{s:ff}

In this section, we summarize the connection between charge conservation and force-free equilibrium.
The closed magnetosphere of a magnetar and, to a lesser extent, the open circuit of a rotating neutron star,
has an enormous self-inductance:  the energy of the charges supporting the background current is miniscule
compared with the energy stored in the non-potential magnetic field.   The magnetic field is assumed to be weakly
sheared, meaning that a strong `guide' field is present, to be identified with the mean helical field
encountered in Section \ref{s:toroidal}.  The force balance separates into a transverse component,
 which is dominated by the electromagnetic field, and the longitudinal dynamics of the
 charges, which is not force free.  Our focus here is on the slow drift of charges perpendicular
 to the guide field.

Here and in the remainder of this paper, we work in cartesian geometry.   The guide magnetic field flows in the $z$ direction;
hence
\be
   {\bf B}(x,y)  = B_z\hat z + {\bf B}_\perp = B_z\hat z  + \bnabla A_z\times \hat z,
\ee
where ${\bf A}$ is the vector potential.
The symbol $\perp$ represents a projection onto the $x-y$ plane.  In this section, our focus is on
the motion of charges perpendicular to the guide field, and so we only need the perpendicular electric field ${\bf E}_\perp$.
The magnetic field is assumed to deform slowly (e.g. $E^2 \ll B^2$), as will be justified after the fact for
the linear modes uncovered in this paper.

To an excellent approximation,
the charges (of sign $\pm q$) are guided along the magnetic field, with a finite mean speed
$\beta^\pm\simeq \beta_z^\pm$ that may depend on the sign of $q$.  
The transverse drift speed $\bbeta_\perp^\pm$ is easily obtained from the Lorentz force equation,
\be
\gamma mc \left({\partial\bbeta_\perp^\pm\over\partial t}
+ \beta_z^\pm{\partial\bbeta_\perp^\pm\over\partial z}\right)
= \pm q\left({\bf E}_\perp + \bbeta_\perp^\pm\times{\bf B}_z + \bbeta_z^\pm\times{\bf B}_\perp\right),
\ee
when $|\partial^2\bbeta_\perp^\pm/\partial t^2|$,
$c^2|\partial^2\bbeta_\perp^\pm/\partial z^2| \ll \omega_c^2 \bbeta_\perp^\pm$, where $\omega_c = qB_z/\gamma mc$:
\be\label{eq:betaperp}
\bbeta_\perp^\pm \simeq {{\bf E}_\perp\times{\bf B}_z\over B_z^2} + {\beta_z^\pm\over B_z}{\bf B}_\perp
     + {\rm O}\left({1\over \omega_c t}\right).
%    \pm {1\over \omega_c B_z}\left({\partial{\bf E}_\perp\over\partial t} +
%      \bbeta_z^\pm\times{\partial{\bf B}_\perp\over\partial t}\right).
\ee
Taking the perpendicular divergence of this, combining with the equation of charge conservation,
$\bnabla_\perp\cdot (\rho^\pm \bbeta_\perp^\pm) = - \partial(\rho^\pm\beta_z^\pm)/\partial z
- c^{-1}\partial\rho^\pm/\partial t$, and neglecting the terms proportional to $\omega_c^{-1}$, gives
\be\label{eq:charcon}
   {\partial\over\partial t}\left({\rho\over B_z}\right) + c\bbeta_{E\times B}\cdot\bnabla_\perp\left({\rho\over B_z}\right)
+ {{\bf B}_\perp\cdot\bnabla_\perp\over B_z}\left({J_z\over B_z}\right) + {\partial\over\partial z}\left({J_z\over B_z}\right)
= 0.
\ee
Here $\rho^\pm$ is the density of positive (negative) charge, $\rho = \rho^+ + \rho^-$,
$J_z = (\rho^+ \beta_z^+ + \rho^-\beta_z^-) c$, and $\bbeta_{E\times B}
\equiv {\bf E}_\perp\times{\bf B}_z/B_z^2$.

The quantity explicitly conserved here is the charge flow guided
along a magnetic flux tube, as may be seen by multiplying Equation (\ref{eq:charcon}) by a small constant flux
$\delta \Phi \simeq \delta A_\perp \cdot B_z$, and defining the linear charge density $\delta \lambda =
\delta A_\perp\rho$ and current $\delta I_z = J_z \delta A_\perp$:
\be\label{eq:charcon2}
   {\partial\delta\lambda\over\partial t} + c\bbeta_{E\times B}\cdot\bnabla_\perp\delta\lambda 
+ {{\bf B}_\perp\cdot\bnabla_\perp \delta I_z\over B_z} + {\partial\delta I_z\over\partial z} = 0.
\ee

The tiny effects of particle inertia can be included, in a manner analogous to the treatment of non-relativistic
magnetofluids, by taking the transverse curl of the Euler equation
(now with $\partial/\partial z \rightarrow 0$),
\be
\gamma mnc {\partial\bbeta_\perp\over\partial t} = \rho{\bf E}_\perp + {1\over c}{\bf J}_\perp\times{\bf B}_z
+ {1\over c}{\bf J}_z\times{\bf B}_\perp.
\ee
Here, $n = n^+ + n^- = \rho^+/q - \rho^-/q$ and the transverse momentum density is
$(\gamma mc)n\bbeta_\perp = \gamma mc(n^+\bbeta^+_\perp + n^-\bbeta^-_\perp)$.
Defining the vorticity ${\bf w}_z = w\hat z = \bnabla_\perp\times(\bbeta_\perp c)$,
the right-hand side of Equation (\ref{eq:charcon}) is replaced with
\be\label{eq:inertia}
   {\gamma mnc\over B_z^2} {\partial w\over\partial t} = 
   {\gamma mnc^2\over B_z^3} {\partial\over\partial t}\left(\nabla_\perp^2\phi - \beta_z\nabla_\perp^2 A_z +
   \bnabla_\perp A_z\cdot\bnabla_\perp\beta_z\right),
   \ee
which is suppressed compared with the left-hand side by a factor $\sim 4\pi\gamma mnc^2/B_z^2$.  In the case of
a pulsar, the suppression is almost complete ($\sim 10^{-15}-10^{-20}$).

The leading term on the right-hand side of Equation (\ref{eq:inertia}) will be familiar from the
standard treatment of tearing in a non-relativistic, incompressible magnetofluid (e.g. \citealt{biskamp05}).  In the
non-relativistic regime, the transverse velocity potential $\varpi$ is proportional to the electrostatic potential
that describes ${\bf E}\times{\bf B}$ drift:  ${\bf E} = -\bnabla\phi$ and ${\bf v}_{E\times B} = \hat z \times \bnabla\varpi$
imply $\varpi = c\phi/B_z$ and $w = \nabla^2\varpi = (c/B_z)\nabla^2\phi$.  In the regime of extreme magnetization, the
electromagnetic field plays a central role; but we will see that a kinetic description of linear
perturbations still is written most succinctly in terms of an effective Langrangian displacement field (Section \ref{s:kinetic}).

\section{Kinetic Description of Linear Perturbations}\label{s:kinetic}

The unperturbed state contains a non-potential
magnetic field that varies in cartesian direction $x$ but not in $y$ or $z$,
\ba\label{eq:B0}
   {\bf B}_{\perp 0}(x) &=& B_{y0}(x)\hat y = -A_{z0}'(x)\hat y\nn
   B_{z0} &=& \sqrt{{\cal B}_{\parallel0}^2 - B_{y 0}^2} \simeq {\cal B}_{\parallel0} - {B_{y0}^2\over 2{\cal B}_{\parallel0}};
   \quad\quad{\cal B}_{\parallel0}={\rm const}.
\ea
The second equation describes force-free equilibrium, corresponding to $(d/dx)(B_{y0}^2 + B_{z0}^2) = 0$.
The example investigated in detail below is a periodic shear with a wavelength $2\pi/k_x$ that can be adjusted with
respect to the skin depth:
\be\label{eq:back}
B_{y0}(x) = {\cal B}_{\perp0}\sin(k_xx);  \quad \quad A_{z0}(x) = -{{\cal B}_{\perp0}\over k_x}\cos(k_x x);  \quad\quad J_{z0}(x) = {c\over 4\pi}k_x {\cal B}_{\perp0}\cos(k_xx);\quad\quad {\cal B}_{\perp0}={\rm const}.
\ee

The perturbed system remains uniform in $z$.
The electromagnetic field is the sum of the background (\ref{eq:B0}) and a time-dependent
perturbation (labelled 1) that depends on $x$, $y$ and $t$,
\ba\label{eq:fields}
   {\bf B}_\perp &=& {\bf B}_{\perp0} + \bnabla A_{z1}(x,y)\times \hat z; \nn
   {\bf E}_\perp &=& -\bnabla_\perp \phi_1;\quad
   E_\parallel \equiv {\bf E}\cdot\hat B \simeq -{1\over c}{\partial A_{z1}\over\partial t} -
    {B_{y0}\over B_z}{\partial\phi_1 \over \partial y}.
\ea
The guide field $B_z$ is taken effectively to be a constant, $B_z \simeq {\cal B}_{\parallel 0}
\gg {\cal B}_{\perp0}$.

 In a quantizing magnetic field,
  the unperturbed particle distribution function $f_0$ may depend on fewer than three momentum coordinates.
  For example, if the current sheet is translationally symmetric in one coordinate ($y$), as is assumed here,
  then there is no ${\bf E}\times{\bf B}$ drift out of the plane of the sheet in the background state.   
  In this situation, the perturbed Boltzman equation may involve derivatives
  with respect to all three spatial coordinates,\footnote{Although the modes considered here are invariant in $z$,
    the higher-frequency modes examined in Paper II are not:  they are a generalization of the shear Alfv\'en wave with
    $k_z \sim k_{p0}$.  but only derivatives with respect to the momentum $p$ parallel
  to the magnetic field.  Furthermore, the magnetic field is weakly sheared, meaning that
  the background distribution function depends on $x$ and $p \simeq p_z$.
  We write}
\be
f(x,y,p,t) = f_0(x,p) + f_1(x,y,p,t)
\ee
and, perturbing Equation (\ref{eq:betaperp}),
\be
\bbeta = \bbeta_0 + \bbeta_1 \simeq \beta_0\left(\hat z + {{\bf B}_{\perp 0}\over B_z}\right) +
       {{\bf E}_{\perp 1}\times {\bf B}_z\over B_z^2} + \beta_0{{\bf B}_{\perp 1}\over B_z} + \beta_1{{\bf B}_{\perp 0}\over B_z}.
\ee
    An electric field is excited in the plane of the current sheet
     in the perturbed state, so that $\beta_{x,1} \neq 0$ and one must include a term
$\bbeta_{\perp 1}\cdot\bnabla f_0 = \beta_{x,1}\partial f_0/\partial x$ in the Boltzman equation to represent the advection of charges across the magnetic field.
       
The distribution function is taken to be a narrow top hat centered at momentum $\bar p_0 \equiv \bar\gamma_0\bar\beta_0 mc$ for $q > 0$ (and
$-\bar p_0$ for $q < 0)$.  When there is a mixture of a positive (negative) charges, each with space density $n_0^+$ ($n_0^- = n_0 - n_0^+$), 
\ba
f_0^+(p) &=& {n_0^+\over \Delta p_0}\Theta(p-p_{0-})\Theta(p_{0+}-p); \nn
f_0^-(p) &=& {n_0^-\over \Delta p_0}\Theta(p+p_{0+})\Theta(-p_{0-}-p),
\ea
where $p_{0\pm} = \bar p_0 \pm \Delta p_0/2$, and $\Theta$ is the Heaviside function.\footnote{$\Theta(x) = 0$ $(1)$ for $x<0$ ($x>0$).}
We also assume that $1-\bar\beta_0 \gg (B_{y0}/B_{z0})^2$, corresponding to $\bar\gamma_0 \ll B_{z0}/{\cal B}_{\perp0}$.
The background particle density is
\be
n_0(x) = n_0^+(x) + n_0^-(x) = {1\over 4\pi \bar\beta_0q}B_{y0}'(x),
\ee
and the skin depth
\be
k_{p0}^2 = {4\pi n_0 q^2\over \bar\gamma_0^3mc^2} = {1\over\bar\beta_0\bar\gamma_0^3} {qB_{y0}'(x)\over mc^2}.
\ee
The strong momentum dependence of $k_{p0}$ is a consequence of the narrow momentum distribution assumed;
more generally $k_{p0} \propto \langle \gamma_0^{-3}\rangle^{1/2}$.

The background is invariant under translations in $y$, and so each mode is fourier decomposed as
\be\label{eq:fourier}
A_{z1} = \widetilde A_{z1}(x) e^{(s-i\omega)t + ik_yy}; \quad f_1 = \widetilde f_1(x) e^{(s-i\omega)t + ik_yy}.
\ee
Our goal is to calculate the mode growth rate $s$ and real frequency $\omega$ as functions of $k_{p0}/k_x$, $k_y/k_x$.
Although $\omega$ vanishes in the charge-symmetric state (counter-streaming $e^+$ and $e^-$ with vanishing net charge density,
$n_0^+ = n_0^-$), we find that growing modes are overstable in a charge-asymmetric state.   This will be be important in the
application to pulsars, in particular with regard to the phenomenon of sub-pulse drift.

The perturbed Boltzmann equation reads
\be\label{eq:Boltzperp}
   {\partial f_1^\pm\over\partial t} + c\bbeta_0^\pm\cdot\bnabla f_1^\pm + c\bbeta_1^\pm\cdot\bnabla f_0^\pm =
   \mp q({\bf E}_1\cdot\hat B){\partial f_0^\pm\over \partial p},
\ee
where $\pm$ labels positive and negative charges.
Substituting Equation (\ref{eq:betaperp}) and integrating the sum $q(f_1^+ - f_1^-)$ over $p$ gives
\be
   {\partial\rho_1\over\partial t} + c\bbeta_{E\times B}\cdot\bnabla\rho_0 + {{\bf B}_{\perp 0}\over B_{z0}}\cdot\bnabla J_{z1} 
   + {{\bf B}_{\perp 1}\over B_{z0}}\cdot\bnabla J_{z0}  = 0,
\ee
consistent with Equation (\ref{eq:charcon}).
Here the perturbed charge and current densities $\rho_1$, $J_{z1}$ are
\be\label{eq:rho1}
\rho_1 = q \int dp (f_1^+ - f_1^-);   \quad\quad J_{z1} = q \int dp v (f_1^+ - f_1^-).
\ee

In regions where the current distribution evolves slowly, one may take $\partial/\partial t\rightarrow 0$
and Fourier transform in $y$ to get a modified Grad-Shafronov equation,
\be\label{eq:GS}
c\phi_1 {d\rho_0\over dx} = A_{z1}{dJ_{z0}\over dx} + B_{y0} J_{z1}.
\ee
The electrostatic term on the left-hand side generalizes the equation describing the ideal flux perturbation in
the standard tearing mode analysis.

Returning to the fully time-dependent problem, we adopt a useful shorthand
\be
\hat B\cdot\bnabla \rightarrow ik_y {B_{y0}(x)\over B_{z0}} \rightarrow ik_\parallel(x).
 \ee
Then Equation (\ref{eq:Boltzperp}) reads
%\be
%\left\{s + i[c\bbeta_0\cdot(k_y\hat y)-\omega]\right\}f_1^\pm +
%c\bbeta_1^\pm\cdot\left({\partial f_0\over\partial x}\hat x\right) = \mp q{\bf E}_1\cdot\left({\partial f_0\over\partial p}\hat z\right).
%\ee
%Further subsituting Equation (\ref{eq:betaperp}) (to leading order in $s/\omega_c$) reproduces 
\be\label{eq:f1}
f_1^\pm = {ik_y c\over s'_\pm B_{z0}} \left(\phi_1 - \bar\beta_0A_{z1}\right){\partial f_0^\pm\over\partial x}
\pm {q\over s'_\pm}\left[(s-i\omega)A_{z1} + ik_\parallel c\phi_1\right]{\partial f_0^\pm\over\partial p},
\ee
where
\be
s'_\pm \equiv s + i(\pm\bar\beta_0 ck_\parallel - \omega).
\ee
Integrating the quantity $q(s'_+f_1^+ - s'_-f_1^-)$ over $p$ gives a modified conservation equation for
the longitudinal current,
\be
(s-i\omega)\rho_1 + ik_\parallel J_{z1} = ik_yc\left({\phi_1\over B_{z0}}{d\rho_0\over dx} - {A_{z1}\over B_{z0}}{dJ_{z0}\over dx}\right).
\ee
This reduces to Equation (\ref{eq:GS}) as $s - i\omega \rightarrow 0$.
The perturbed current is conserved along each magnetic flux tube only in the absence of transverse gradients in $\rho_0$ or $J_{x0}$.

\subsection{Perturbed Charge and Current Density}\label{s:charge}

Here we evaluate the perturbed charge and longitudinal current densities, as given by Equation
(\ref{eq:rho1}), using Equation (\ref{eq:f1}) for $f_1^\pm$.  When evaluating
the transverse gradient $\partial f_0^\pm/\partial x$, we choose $n_0^\pm$ to be constant (or, more precisely, assume a
gradient scale much exceeding $2\pi/k_x$).  The $p$-integral of the term proportional to $\partial f_0^\pm/\partial x$
is straightforward due to the narrowness of the momentum distribution, giving
\be
4\pi\rho_1 = i\left(k_{p0}^2-{4\pi \over c B_{y0}}{dJ_{z0}\over dx}\right){ck_\parallel\over s'^2}
\left[(s-i\omega)A_{z1} + ick_\parallel\phi_1\right];
\ee
\be
{4\pi\over c}J_{z1} = {4\pi \over cB_{y0}}{dJ_{z0}\over dx}\left\{
     \left[{(s-i\omega)^2\over s'^2} - 1\right]A_{z1}  + i{ck_\parallel(s-i\omega)\over s'^2}\phi_1\right\}
- k_{p0}^2 {s-i\omega \over s'^2} \left[(s-i\omega)A_{z1} + ic k_\parallel \phi_1\right],
\ee
where
\be\label{eq:sprimeinv}
   {n_0\over s'^2} \equiv {n_0^+\over [s + i(\bar\beta_0ck_\parallel-\omega)]^2} + {n_0^-\over [s + i(-\bar\beta_0ck_\parallel-\omega)]^2}.
\ee
In the case of a neutral plasma, the mode is purely growing or decaying ($\omega = 0$) and this reduces to
\be\label{eq:sneut}
   {1\over s'^2} = {s^2 - (\bar\beta_0 ck_\parallel)^2\over [s^2 + (\bar\beta_0 ck_\parallel)^2]^2}\quad\quad(n_0^+ = n_0^-).
\ee
In what follows, we take $\bar\beta_0$ to be independent of $x$, that is, we assume that the high-wavenumber magnetic flux
component driving reconnection is superposed on a smoother component that contributes a greater fraction of the
total current (Section \ref{s:toroidal}).

These equations for $\rho_1$ and $J_{z1}$ can be further simplified by a change of variables.  The presence of a strong guide
field allows the definition a hydromagnetic displacement field $\bxi_\perp = (\xi_x, \xi_y)$ and a local `rest' frame in
which ${\bf E}_\perp = -c^{-1}(\partial\bxi_\perp/\partial t)\times{\bf B}_z \rightarrow 0$.   Then one may write
\be\label{eq:hydro}
\phi_1 \rightarrow {i(s-i\omega)B_{z0}\over ck_y}\xi_x; \quad\quad
A_{z1} \rightarrow B_{y0}\xi_x + A_{z1}^{\rm rec} = B_{y0}(\xi_x + \xi_x^{\rm rec}).
\ee
The first term in the expression for $A_{z1}$ represents the effect of a hydromagnetic displacement.  This is the
magnetic perturbation associated with the second term in $E_\parallel$ in Equation (\ref{eq:fields}).
This displacement scales as $\xi_x \propto x$  near $x=0$, whereas the magnetic potential perturbation $A_{z1}(0)$ remains finite
in a tearing instability.   Therefore, the non-ideal displacement $\xi_x^{\rm rec}$ diverges as $\sim x^{-1}$
near $x = 0$, where $B_{y0} \rightarrow 0$.  The linear combination
\be
(s-i\omega)A_{z1} + ick_\parallel\phi_1 \rightarrow (s-i\omega)B_{y0}\xi_x^{\rm rec} \;\sim\; x^0
\ee
vanishes only in the ideal regime.

To evaluate the mode spectrum, we must iteratively compute the profiles of
$A_{z1}$ and $\phi_1$, or equivalently of $\xi_x$ and $\xi_x^{\rm rec}$.   Coulomb's law implies
\be
-{\partial^2\phi_1\over\partial x^2} + k_y^2\phi_1 = 4\pi \rho_1
= i {k_\parallel c(s-i\omega)\over s'^2}\left(k_{p0}^2-{1\over B_{y0}}{d^2B_{y0}\over dx^2}\right)A_{z1}^{\rm rec}.
\ee
Substituting Equation (\ref{eq:hydro}) gives\footnote{In what follows,
we discard the $\;\widetilde\null\;$ from functions of $x$ appearing in the Fourier decomposition (\ref{eq:fourier}).}
\be\label{eq:Coul}
{d^2\xi_x\over dx^2} = k_y^2\xi_x -
{(ck_\parallel)^2\over s'^2}\left(k_{p0}^2-{1\over B_{y0}}{d^2B_{y0}\over dx^2}\right)\xi_x^{\rm rec}
\ee
Similarly, Ampere's law $-\partial^2A_{z1}/\partial x^2 + k_y^2A_{z1} = (4\pi/c)J_{z1}$ implies
\be\label{eq:Amp}
   {d^2\over dx^2}[B_{y0}(\xi_x+\xi_x^{\rm rec})] = \left(k_y^2B_{y0} + {d^2B_{y0}\over dx^2}\right)(\xi_x+\xi_x^{\rm rec})
  + {(s-i\omega)^2\over s'^2}\left(k_{p0}^2B_{y0} - {d^2B_{y0}\over dx^2}\right)\xi_x^{\rm rec}.
\ee
These equations simplify further when the current profile is harmonic in coordinate $x$.   One sees that the coupling
between the electrostatic and vector potential perturbations is driven by the non-ideal part of the displacement field.

\section{Neutral Pair Plasma with a Strong Guide Magnetic Field}\label{s:neutral}

This section is devoted to an evaluation of the mode spectrum,
\be\label{eq:solution}
s = s(k_y/k_x, k_{p0}/k_x), \quad\quad \omega = \omega(k_y/k_x, k_{p0}/k_x),
\ee
in the case of a sinusoidal current profile (Equation (\ref{eq:back})).
This toy problem encapsulates the effects of multiple interacting tearing surfaces,
where the non-potential magnetic field vanishes,
\be\label{eq:surf}
B_{y0} = 0\quad  (x = 0, \pm n\pi).
\ee
The resulting eigenvalue problem is tractable.

We first consider the simplest case of a charge-neutral plasma, $n_0^+ = n_0^-$,
with positive and negative charges counterstreaming, $\beta_{z0}^- = -\beta_{z0}^+$.
Then the tearing modes are purely growing or decaying ($\omega = 0$);  we find
multiple branches at $k_{p0}/k_x \gtrsim 1$.
These results are generalized in Section \ref{s:nonneutral}
to a non-neutral plasma (where $\omega \neq 0$) and in Section \ref{s:single} to a single tearing surface.

\subsection{Eigenvalue Equations}

A simplified form for the eigenvalue equations is found by substituting Equations (\ref{eq:back}) for $B_{y0}$ and
$J_{x0}$ and Equation (\ref{eq:sneut}) for $s'$ in Equations (\ref{eq:Coul}) and (\ref{eq:Amp}), giving
\be\label{eq:Coulneut}
{d^2\xi_x\over d(k_x x)^2} = {k_y^2\over k_x^2}\xi_x -
{(ck_\parallel)^2[s^2-(\bar\beta_0ck_\parallel)^2]\over [s^2 + (\bar\beta_0 ck_\parallel)^2]^2}
\left({k_{p0}^2\over k_x^2} + 1\right)\xi_x^{\rm rec}
\ee
and
\be\label{eq:Ampneut}
   {d^2[B_{y0}(\xi_x + \xi_x^{\rm rec})]\over d(k_xx)^2} = \left({k_y^2\over k_x^2} - 1\right)B_{y0}(\xi_x + \xi_x^{\rm rec}) +
   {s^2[s^2-(\bar\beta_0ck_\parallel)^2]\over[s^2 + (\bar\beta_0ck_\parallel)^2]^2}\left({k_{p0}^2\over k_x^2} + 1\right)B_{y0}\xi_x^{\rm rec},
   \ee
where $k_\parallel = k_yB_{y0}/B_{z0}$.

We search for tearing modes with finite flux perturbation $A_{z1}$ at the surfaces (\ref{eq:surf}), and with
anti-symmetric ${\bf E}\times{\bf B}$ drift: $ \xi_x(x>0) = -\xi_x(x < 0)$.   The flux perturbation is
symmetric, because $B_{y0}$ and $\xi_x$, $\xi_x^{\rm rec}$ both reverse sign across $x = 0$.
Equations (\ref{eq:Coulneut}) and (\ref{eq:Ampneut}) are therefore
solved on the interval $k_x x = [0, \pi/2]$ with the boundary conditions
\be
 {dA_{z1}\over dx}(0) = 0; \quad \phi_1(0) = 0
\ee
at $x = 0$ and
\be\label{eq:bc}
 {dA_{z1}\over dx}(\pi/2) = 0; \quad {d\phi_1\over dx}(\pi/2) = 0
 \ee
 at $k_x x = \pi/2$ (half way to the next tearing surface).   We use a shooting method,
 iterating on $s$ and the slope of $\xi_x$ at $x = 0$ until the boundary conditions (\ref{eq:bc}) are satisfied.

\begin{figure}
  \epsscale{1.1}
  \vskip -0.3in
\plottwo{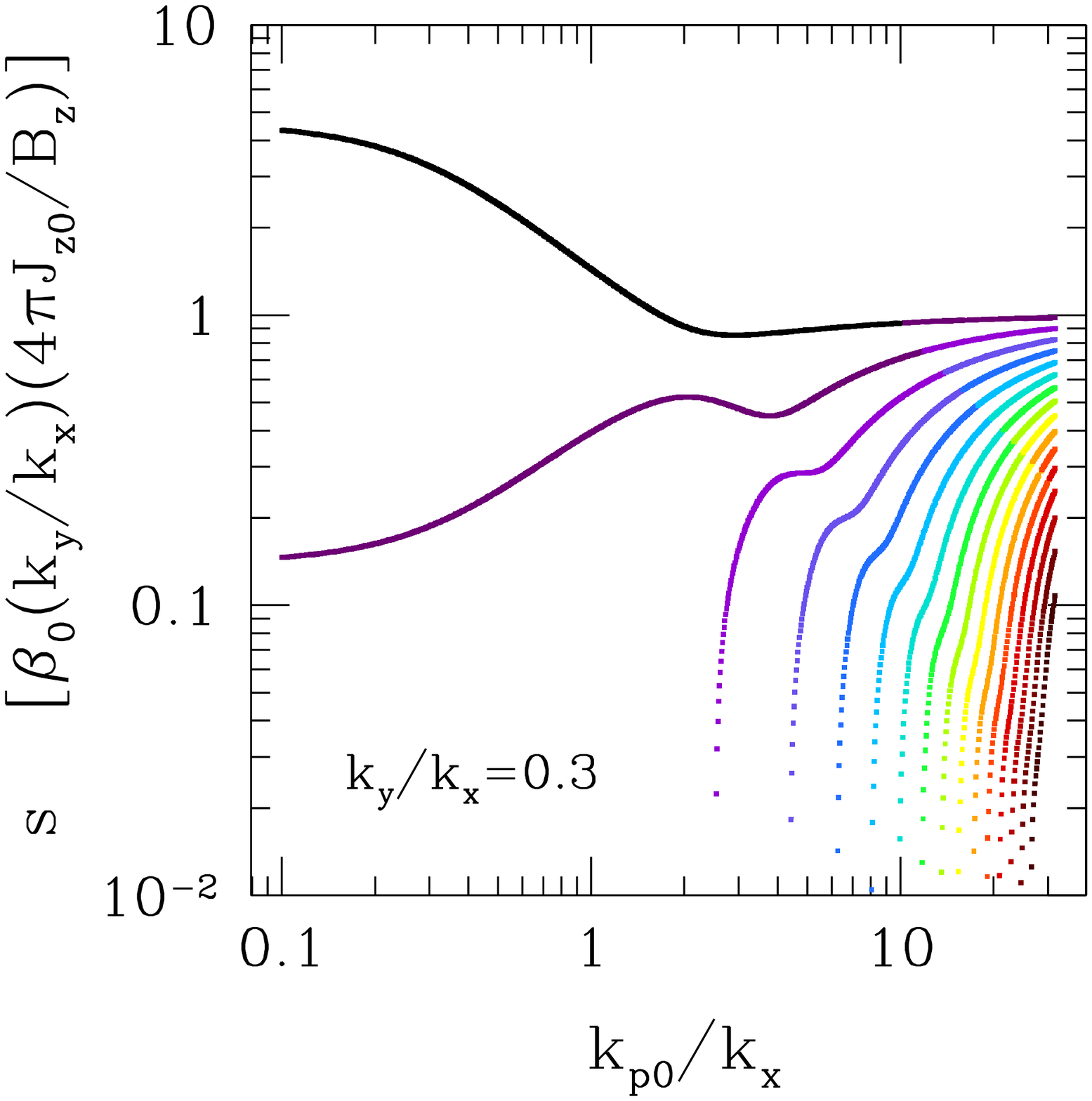}{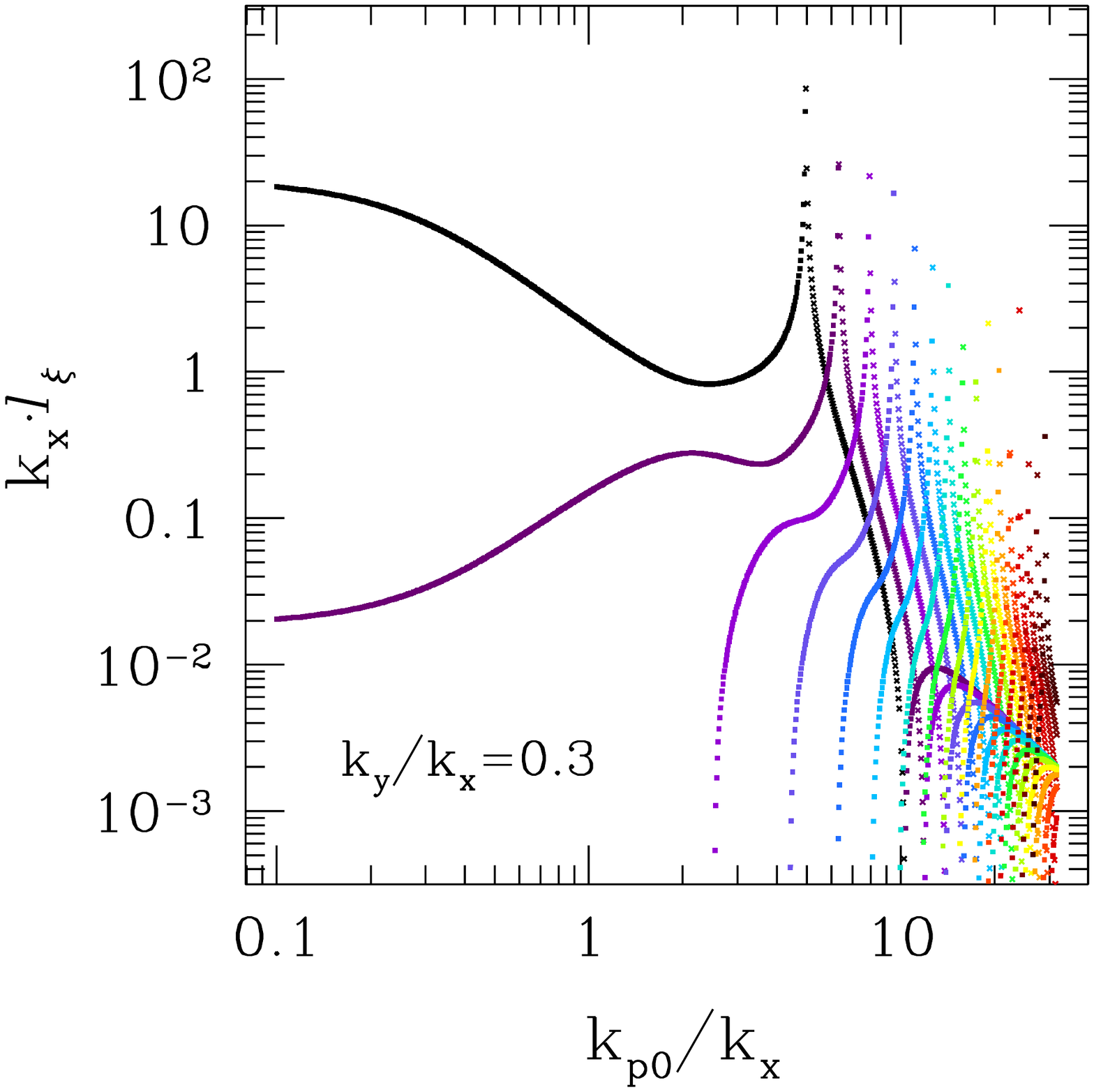}
\vskip -0.8in
\caption{Left panel:  Dependence of mode growth rate $s$ on $k_{p0}/k_x$ in the solution to the eigenvalue
  problem of Section \ref{s:neutral}.  Right panel:  Gradient scale $l_\xi$ of the electrostatic perturbation near $x=0$, defined by
  $\xi_x = (x/l_\xi)A_{z1}(0)/{\cal B}_{\perp0}$.   New branches of the dispersion curve appear as the tearing surfaces move apart relative
  to the skin depth ($k_{p0}/k_x$ increases).   These higher-order modes contain additional nodes in $\xi_x^{\rm rec}$;  their growth rates converge on
  $\widetilde s \sim 1$, suggesting that a flux variation of low $k_x$ triggers a broad spectrum of secondary variations with wavenumber
  ranging from $\sim k_x$ to $\sim k_{p0}$.  Colors indicate the number $N_{\rm rec}$ of nodes in $\xi_x^{\rm rec}$, ranging from $N_{\rm rec} = 0$ (black)
  to $N_{\rm rec} \gtrsim 15$ (dark red).  Dispersion curves show a sudden shift in $N_{\rm rec}$ at high $k_{p0}/k_x$, marking
  a singularity in the electric field at $x = 0$ (where $|l_\xi|^{-1} \rightarrow \infty$ and 
  $E_{x1}(0)$ flips sign).  In the right panel, square points denote $l_\xi > 0$ and crosses $l_\xi < 0$.\label{fig:dispersion}}
\vskip 0.7in
\end{figure}

\begin{figure}
  \epsscale{1.1}
  \vskip -0.3in
\plottwo{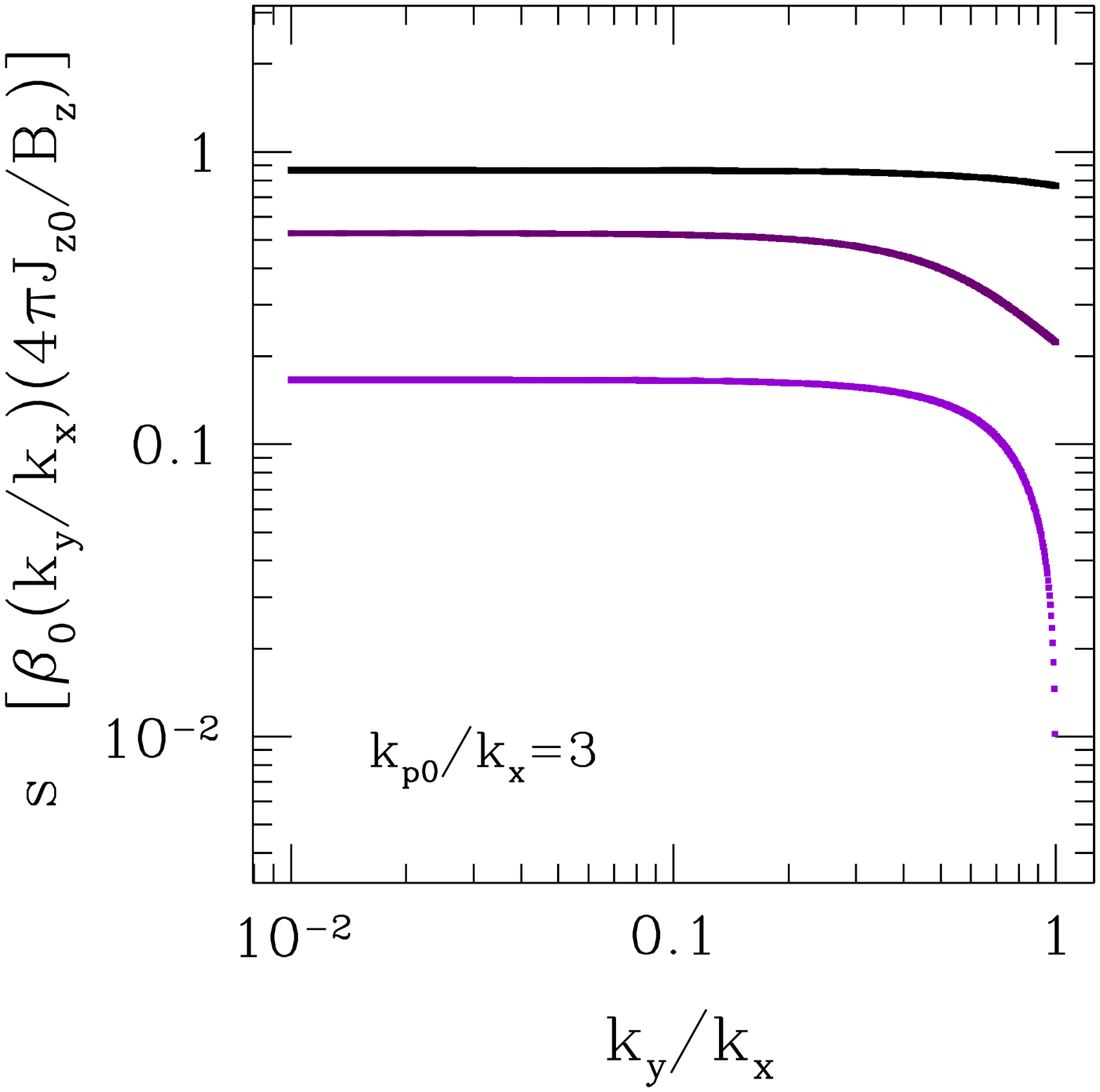}{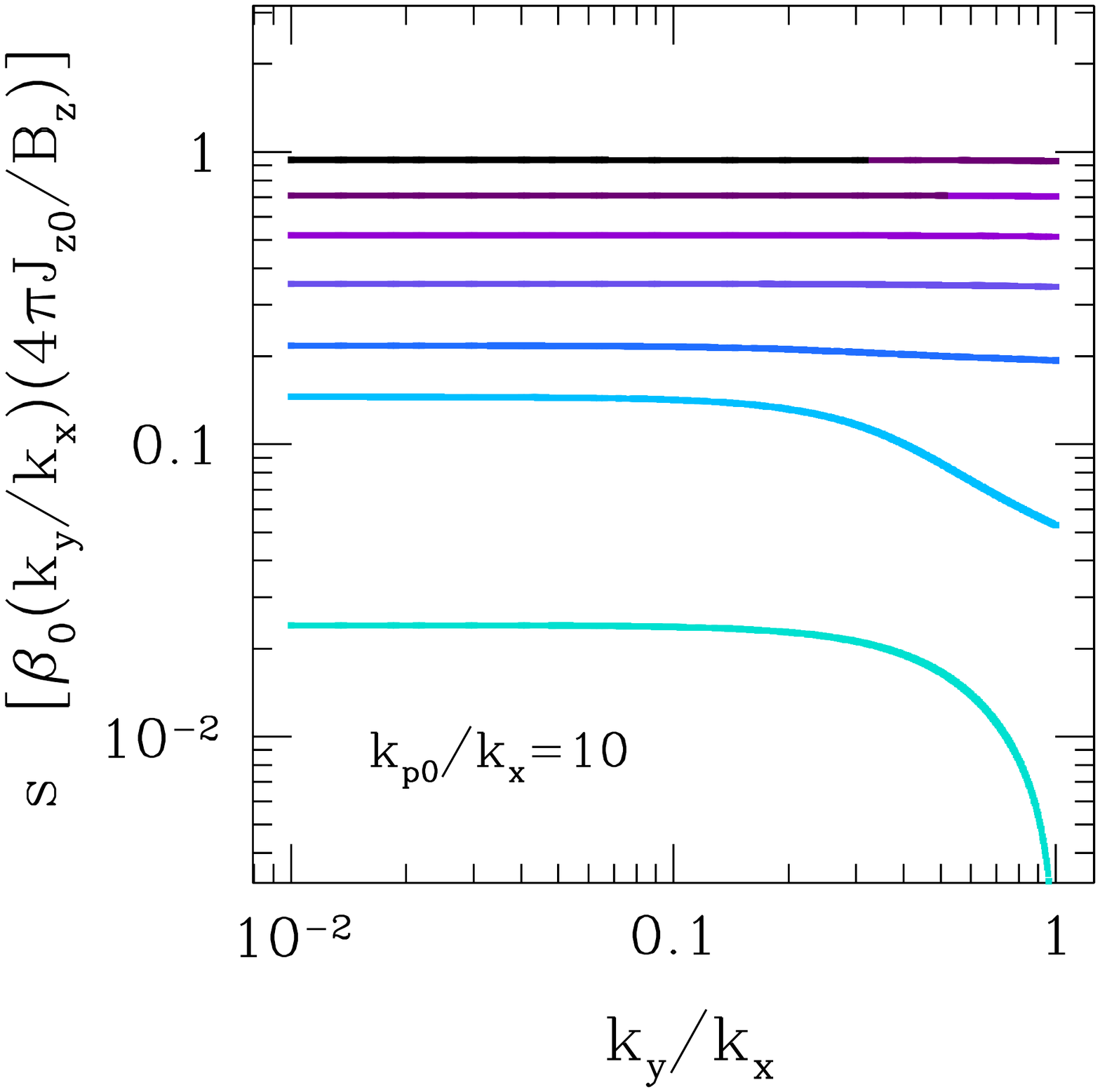}
\vskip -0.8in
\caption{Same as Figure \ref{fig:dispersion}, but now showing the dependence on mode wavenumber $k_y$ at fixed
  ratio of current sheet spacing to skin depth $k_{p0}/k_x$.\label{fig:dispersiony}}
\end{figure}

\begin{figure}
  \epsscale{1.1}
  \vskip -0.3in
  \plottwo{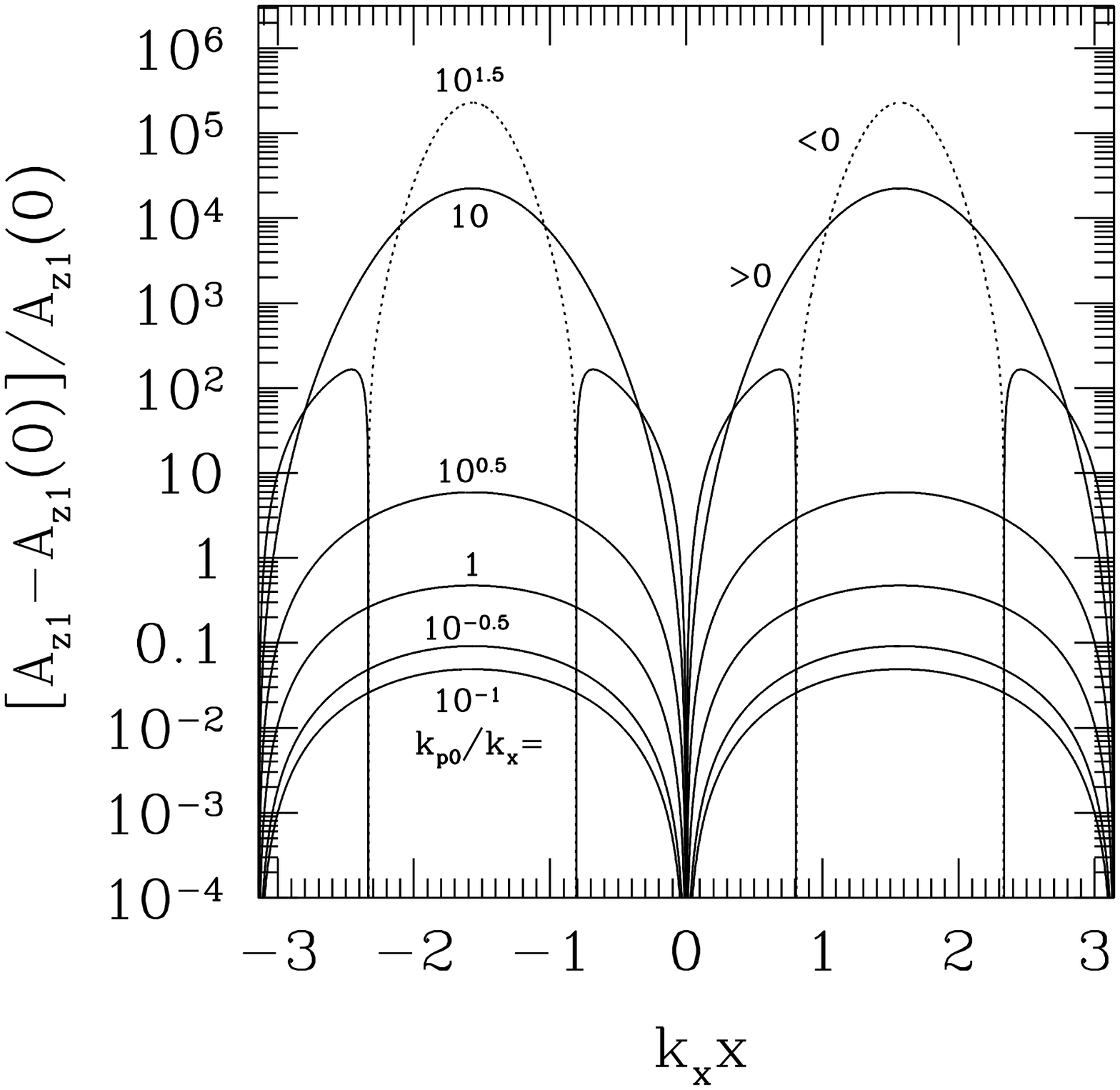}{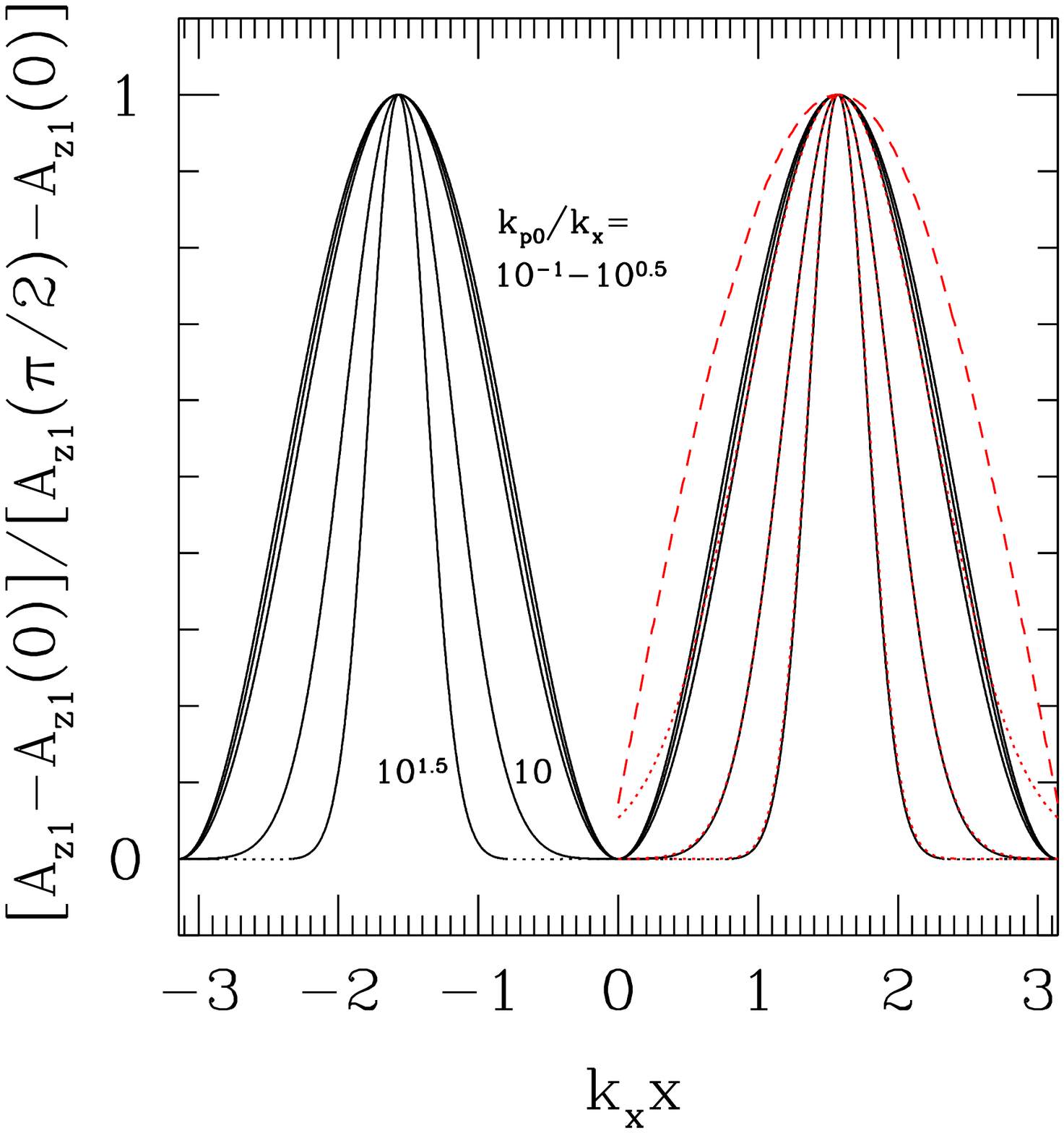}
  \vskip -1.1in
\plottwo{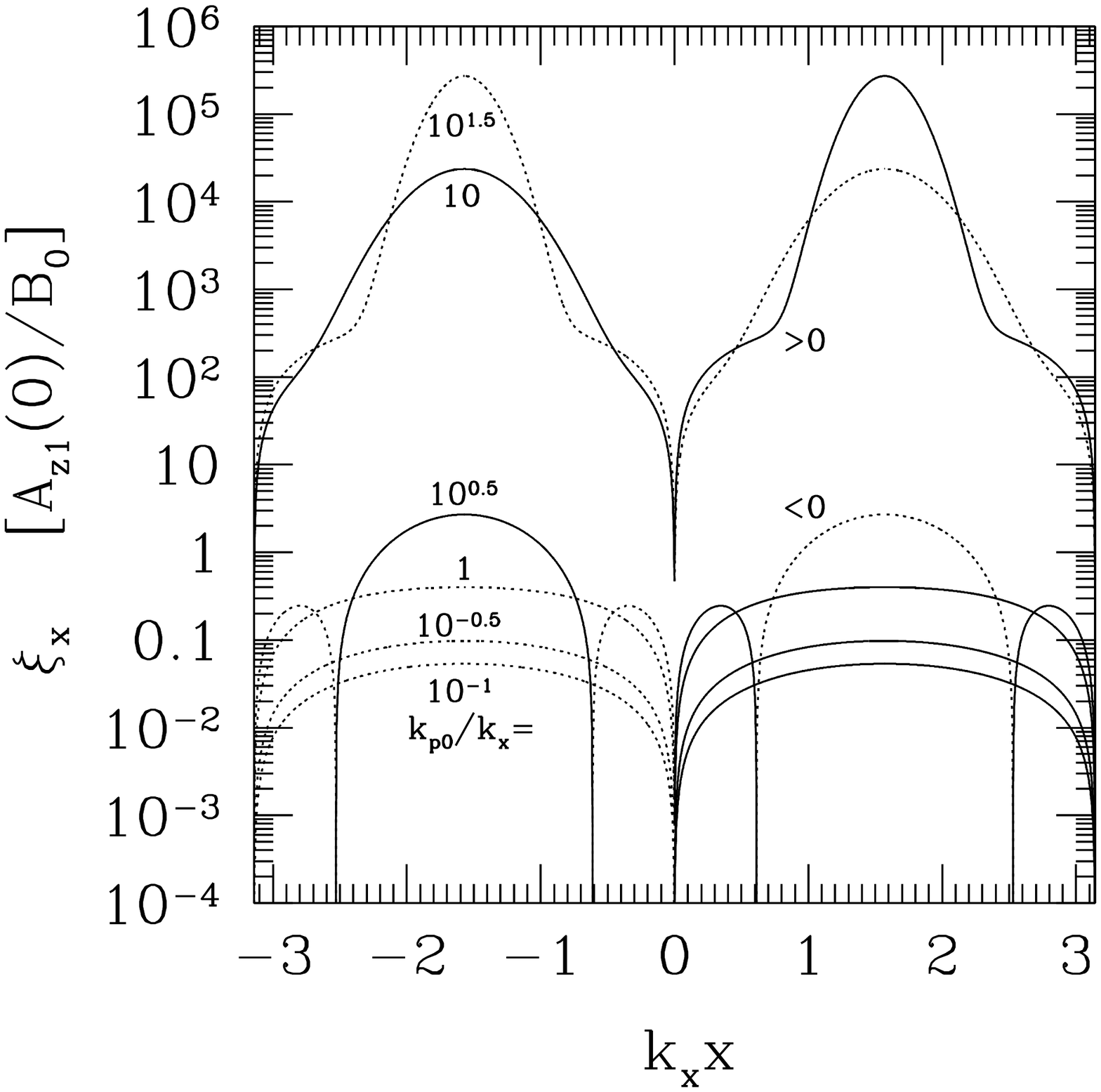}{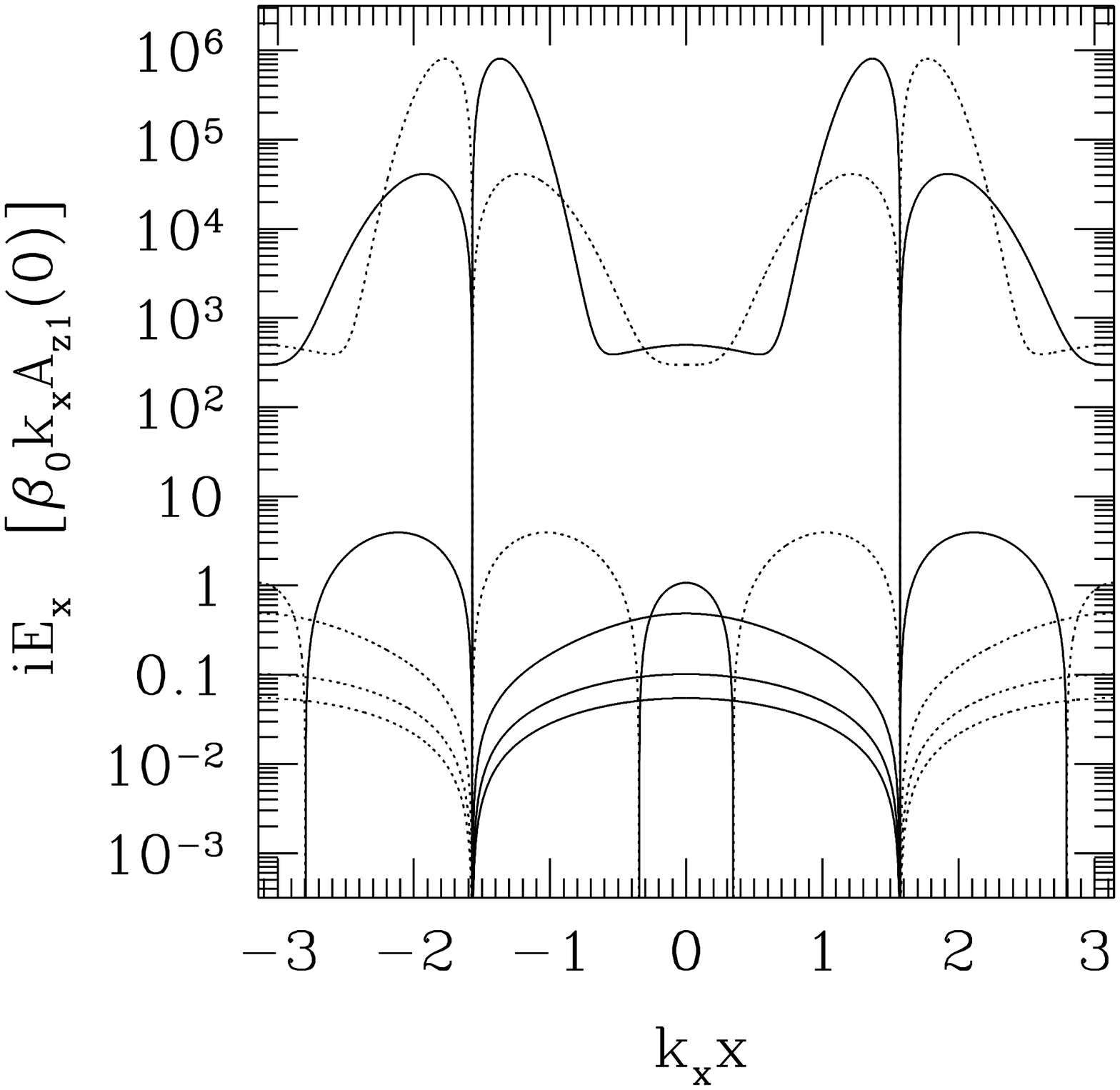}  
\vskip -0.8in
\caption{Top-left panel:  Vector potential perturbation $A_{z1}$, for several ratios $k_{p0}/k_x$ of current sheet spacing to skin depth.
  In all panels, the mode wavenumber $k_y = 0.3k_x$ and dotted curves show perturbations with a reversed sign, e.g. $A_{z1} < 0$
  in the top-left panel.   As $k_{p0}/k_x$ increases, the mode becomes
  strongly concentrated in the bulk, away from the tearing surfaces.    Top-right panel:  $A_{z1}$ but now normalized
  to its peak value at $k_xx = \pm \pi/2$.   Red-dotted lines shown the analytic approximation (\ref{eq:approx2}),
  which applies when $k_{p0} \gtrsim k_x$.  Red-dashed curve shows for comparison the profile (\ref{eq:approx3}) that would obtain if the bulk zone were
  fully in the ideal regime, with $\widetilde s \rightarrow 0$.   Bottom-left panel:  hydromagnetic $x$-displacement $\xi_x$, which is
  proportional to the scalar potential $\phi_1$ (Equation (\ref{eq:hydro})).  Bottom-right panel:  electric field component
  $E_{x1} = -\partial_x\phi_1$ transverse to the $B_y = 0$ surfaces; labelling of curves follows the other panels.
  Note that $A_{z1}$ and $E_x$ are symmetric about $x = 0$, whereas $\xi_x$ and $\phi_1$ are anti-symmetric.\label{fig:mode}}
\vskip 0.3in
\end{figure}

\begin{figure}
  \epsscale{1.1}
  \vskip -0.3in
  \plottwo{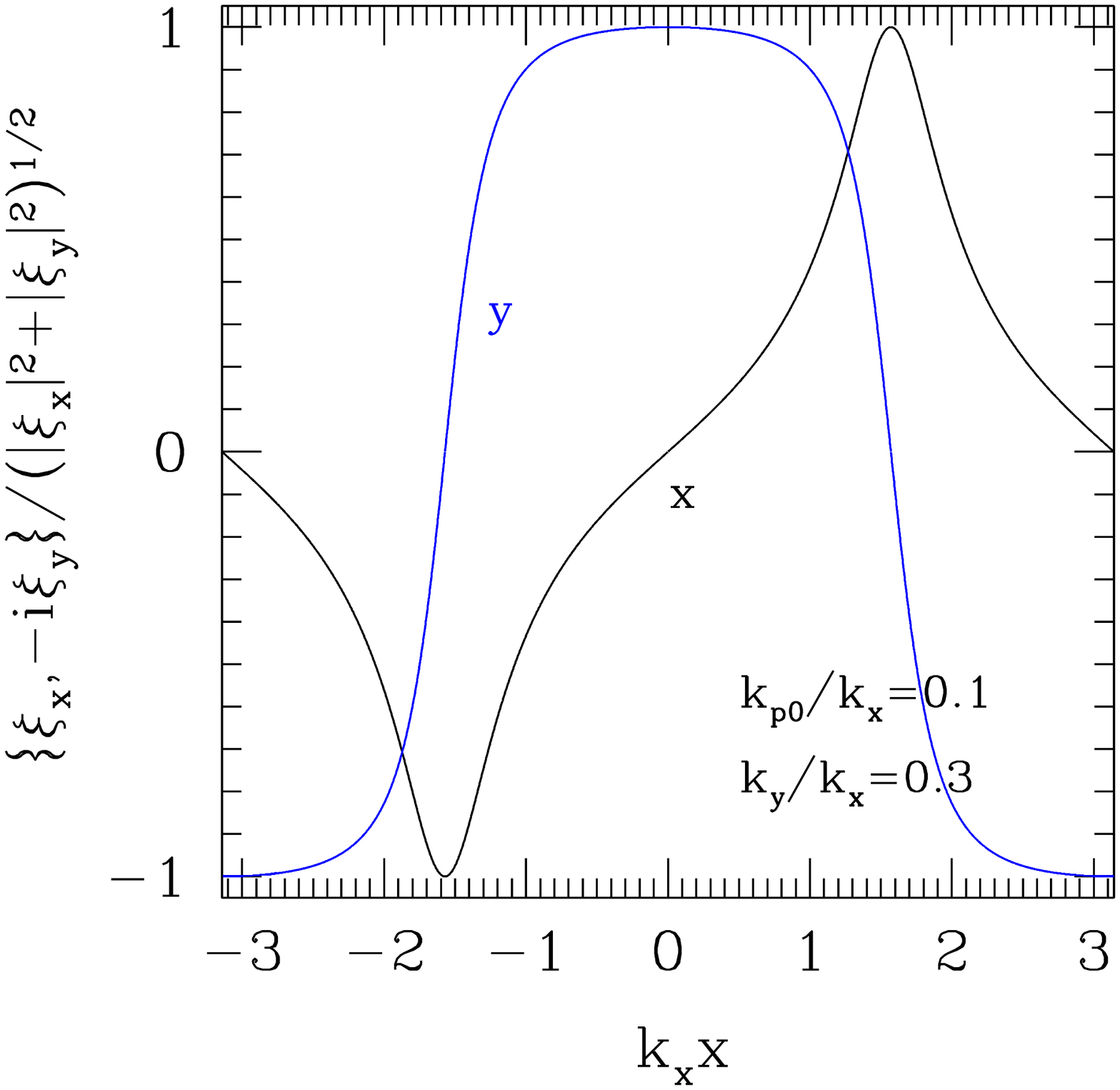}{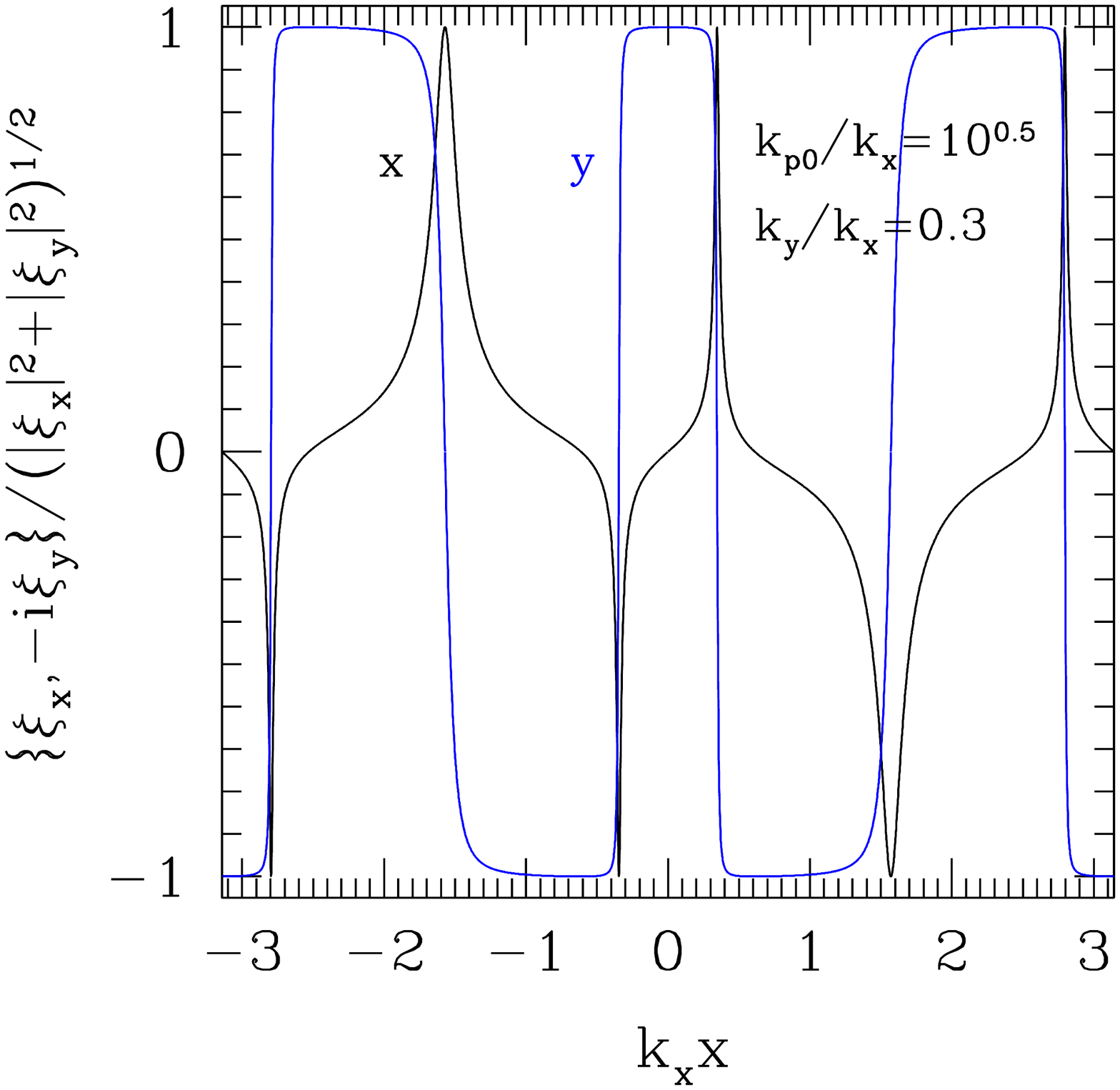}
  \vskip -1.1in
  \epsscale{0.55}
\plotone{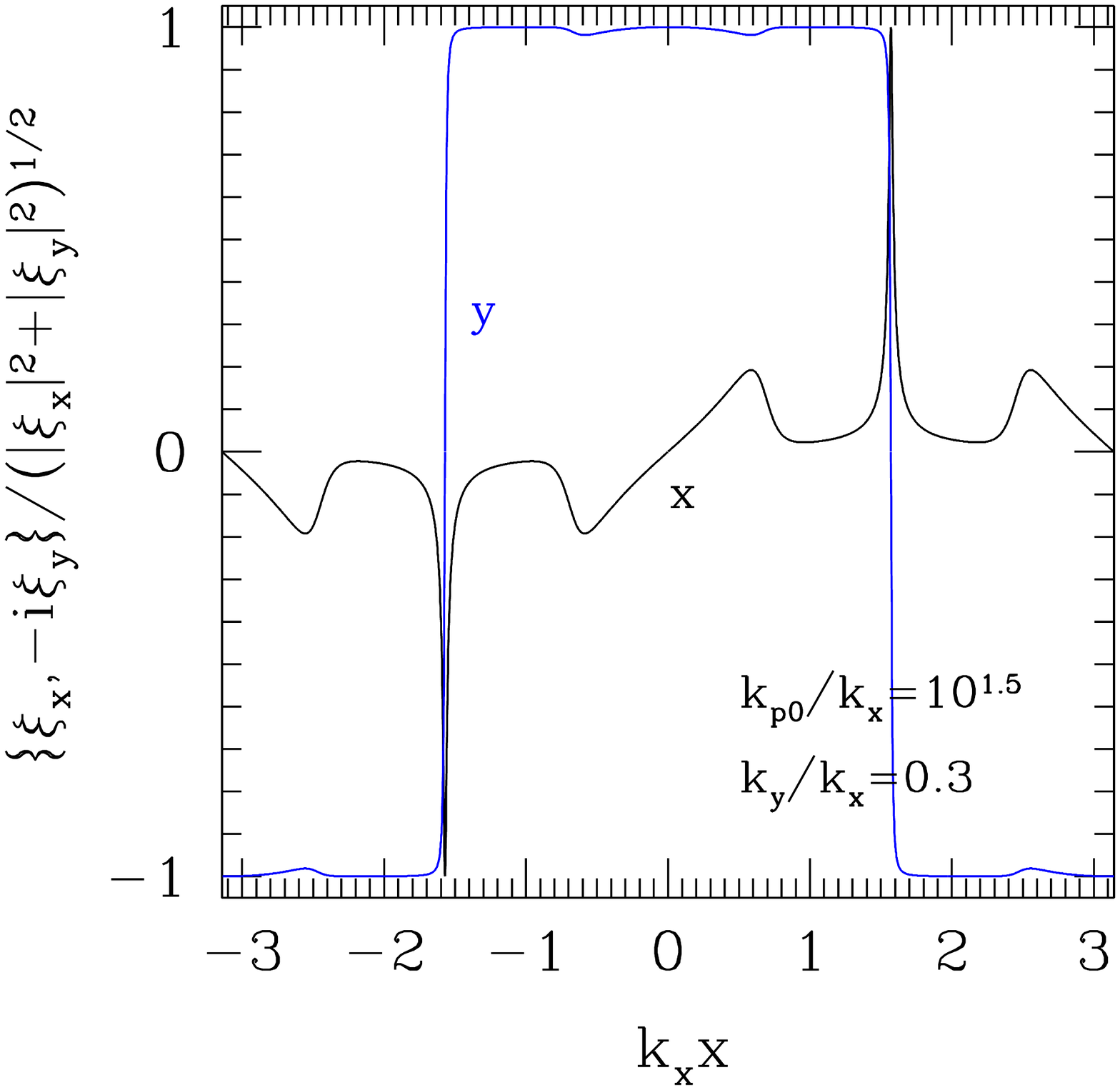}
\vskip -0.8in
\caption{$x$- and $y$-deformations of the strong guide magnetic field, as mediated by
  ${\bf E}\times{\bf B}$ drift, for three of the mode solutions represented in Figure \ref{fig:mode}.  $\xi_x$ always
  vanishes at the tearing surfaces $B_{y0} = 0$ ($x = 0, \pm n\pi$) and $\xi_y$ vanishes at the midpoint between these surfaces.  As the system
  grows in size compared with the skin depth ($k_{p0}/k_x$ increases) the displacement field is increasingly aligned with
  the tearing surfaces ($|\xi_y| \gg |\xi_x|$, excepting at the midpoint.\label{fig:displacement}}
\vskip 0.3in
\end{figure}

\subsection{Growth Rates}

A wide spectrum of modes is present when $k_{p0} \gtrsim k_x$.  Figures \ref{fig:dispersion} and \ref{fig:dispersiony}
respectively show the dependence of growth rate on (i) the ratio $k_{p0}/k_x$ of current sheet spacing
to skin depth and (ii) on the mode wavenumber $k_y$.  The peak growth rate is, when $k_{p0} \gtrsim k_x$,
\be\label{eq:slimit}
s \sim \bar\beta_0 ck_\parallel \sim \bar\beta_0 ck_y {{\cal B}_{\perp0}\over B_{z0}} = \bar\beta_0 {k_y\over k_x}{4\pi J_{z0}\over B_{z0}}.
\ee
In the open pulsar circuit, where $J_{z0} \sim \rho_{\rm co}c = \Omega B_{z0}/2\pi$, the peak growth
rate is therefore comparable to the angular frequency of rotation of the star,
\be\label{eq:slimit2}
s \sim 2\Omega\cdot{k_y\over k_x}.
\ee
Faster growth is seen when the separation between tearing surfaces is smaller than the skin depth, $k_x > k_{p0}$.  

Near a rotating neutron star, the tearing modes uncovered here grow slowly compared with
the Alfv\'en frequency $c/R_{\rm NS}$, where $R_{\rm NS}$ is the stellar radius.  Furthermore, because $s/ck_\parallel$ is never
a large number, no part of the eigenmode can be understand as evolving in the ideal regime.  
The non-ideal displacement $\xi_x^{\rm rec}$ dominates the hydromagnetic displacement $\xi_x$ not only near the surfaces $B_{y0} = 0$
(where $A_{z1}$ is finite and so $\xi_x^{\rm rec} \propto B_{y0}^{-1} \propto x^{-1}$), but everywhere within the bulk.
In other words, these modes cannot be divided into a narrow layer where flux freezing breaks down,
surrounded by a more extended ideal magnetohydrodynamic
zone, as in the standard tearing mode analysis of a collisionless plasma \citep{drake_lee77,basu81,ottaviani93}.

Several branches are apparent in the mode spectrum.  Each branch is characterized by the number $N_{\rm rec}$
nodes\footnote{No similar correlation is found with the number of nodes in the total displacement field $\xi_x$.}
in the profile of the non-ideal displacement field $\xi_x^{\rm rec}$, as defined in Equation (\ref{eq:hydro}),
over the interval $0 < x < \pi/2k_x$.
The mode with the highest growth rate $s$ has a profile with the smallest number of nodes (0).  
An increment in $N_{\rm rec}$ is seen on each branch at a large, discrete value of $k_{p0}$.  Here, the gradient
of $\xi_x$ diverges near $x = 0$, implying a singularity in the transverse electric field -- see the right panels
of Figures \ref{fig:dispersion} and \ref{fig:dispersiony}.

\subsection{Eigenmodes}

The profile of the magnetic potential perturbation $A_{z1}$ is shown in the top two panels of Figure \ref{fig:mode},
for several values of $k_{p0}/k_x$ (ranging from $10^{-1}$ to $10^{1.5}$) and mode wavenumber $k_y = 0.3k_x$.
The main trend is a growing concentration of the mode displacement away from the surfaces $B_{y0} = 0$ as the system
size increases with respect to the skin depth.  For example, when $k_{p0}/k_x = 10^{1.5}$, the magnetic potential
perturbation is $\sim 1\times 10^6$ times larger at the mid-point ($k_xx = \pi/2$) than it is at $x = 0, \pm \pi$.

This result is easily understood quantitatively.  When $k_{p0} \gg k_x$ and
$|\xi_x|$, $|\xi_x^{\rm rec}|$ are large compared with $A_{z1}(0)/{\cal B}_{\perp0}$, Equations (\ref{eq:Coulneut}) and (\ref{eq:Ampneut}) combine 
to give
\be\label{eq:approx}
   {d^2A_{z1}\over d(k_xx)^2} \simeq \left({k_{p0}^2\over k_x^2} + 1\right)
   {[\widetilde s^2-\sin^2(k_xx)][\widetilde s^2 + \sin^2(k_xx)/\bar\beta_0^2]\over [\widetilde s^2 + \sin^2(k_xx)]^2}A_{z1},
\ee
where $ s = \widetilde s\cdot \bar\beta_0 ck_y({\cal B}_{\perp0}/B_{z0})$.  Writing $A_{z1} = C e^{g(x)}$ and approximating
$\widetilde s$, $\bar\beta_0 \simeq 1$, one has to leading order
$g'(x) = [(k_{p0}/k_x)^2+1]^{1/2}\cos(k_xx)/[1+\sin^2(k_xx)]^{1/2}$ and
\be\label{eq:approx2}
 {A_{z1}(x)\over A_{z1}(\pi/2k_x)} = \left\{{\sin^2(k_xx) + [1+\sin^2(k_xx)]^{1/2}\over 1 + 2^{1/2}}\right\}^{[(k_{p0}/k_x)^2+1]^{1/2}}.
\ee
This formula (shown as dotted curves in the top-right panel of Figure \ref{fig:mode}) provides an excellent fit to the
full solution for $k_{p0}/k_x = 10, 10^{1.5}$, and still a good fit for $k_{p0}/k_x = 10^{0.5}$.

A useful comparison can be made with internal tearing in a collisionless but sub-relativistic plasma \citep{ottaviani93}.   Here
the current system is nearly quasi-static throughout the bulk, corresponding to $\widetilde s \ll 1$.  Then Equation (\ref{eq:Ampneut})
reduces to the equilibrium given by Equation (\ref{eq:GS}), which may be written
\be
   {d^2A_{z1}\over dx^2} = (k_y^2 - k_x^2)A_{z1}.
\ee
The solution which is symmetric about $x = \pi/2k_x$ is 
\be\label{eq:approx3}
    {A_{z1}(x)\over A_{z1}(\pi/2k_x)} = \cos\left(\kappa x - {\pi\kappa\over 2k_x}\right) \quad\quad (k_y < k_x),
\ee
where $\kappa = (k_x^2 - k_y^2)^{1/2}$.  This is plotted as the dashed red line in Figure \ref{fig:mode}, showing that
the mode is concentrated in the bulk, but not as sharply as the non-ideal modes obtained above.

We also obtain a simple derivation of the growth rate $\widetilde s \simeq 1$ when $k_{p0}/k_x > 1$,
\be\label{eq:sapprox}
1 - \widetilde s^2 \sim {4\over (1+\bar\beta_0^{-2})(1+k_{p0}^2/k_x^2)}.
\ee
To see this, note that $A_{z1}$ grows exponentially away from the surface $B_{y0} = 0$
but still satisfies the boundary condition $A_{z1}' = 0$ at $|x| = \pi/2k_x$.
Then $-A_{z1}^{-1}d^2A_{z1}/d(k_xx)^2 = O(1)$ at the mid-point, and Equation (\ref{eq:sapprox})
follows by setting $\sin^2(k_xx)= 1$ in Equation (\ref{eq:approx}).

Other details of the eigenmodes are shown in the bottom two panels of Figure \ref{fig:mode} and Figure
\ref{fig:displacement}:  the $x$- and $y-$displacements $\xi_{x,y}$ and the transverse electric field.
This provides an alternative view of how the displacement is concentrated away from the surfaces
where $B_{y0} = 0$ when these surfaces are well separated compared with the skin depth.

\section{Isolated Current Sheet}\label{s:single}

We now shift to consider tearing at an isolated current sheet in a quantizing magnetic field.  This configuration
supports a more restricted set of tearing modes than does the harmonic current distribution
with multiple tearing surfaces that we analyzed in Section \ref{s:neutral}.
Growing modes are found only when the current sheet is relatively narrow; the overtone modes
found previously for $k_x \ll k_{p0}$ are absent when the mode is constrained to have a finite energy.
We also note that our kinetic analysis does not uncover
a separation between a non-ideal sublayer and a more extended, nearly ideal displacement of the
magnetic field -- in contrast with a previous analysis based on a resistive formulation
of force-free electrodynamics \citep{lyutikov03}.  In particular, the standard approximation 
of a nearly uniform magnetic flux perturbation in a resistive sublayer cannot be made in this
situation.  The flux perturbation retains a significant non-ideal component
outside the current sheet; instability
to tearing cannot be expressed in a simple way in terms of the jump in $\partial\log A_{z1}/\partial x$
across a resistive sublayer (e.g. \citealt{biskamp05}).  We have checked that attempting to enforce constant
$A_{z1}$ nearl $x=0$ does not generate self-consistent, finite-energy tearing modes.

\begin{figure}
  \epsscale{0.55}
  \vskip -0.3in
  \plotone{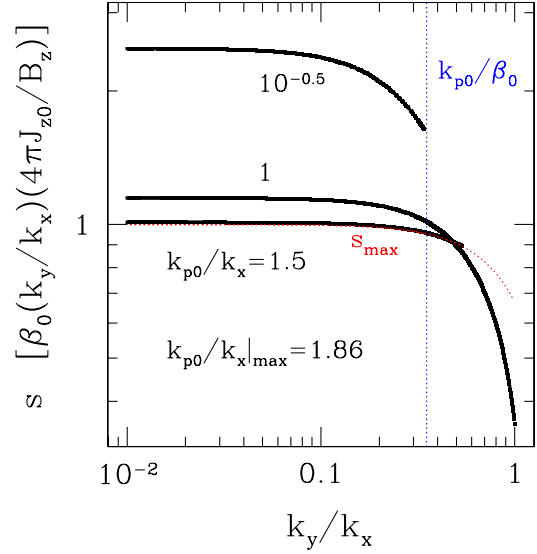}
\vskip -0.8in
\caption{Growth rate of tearing mode at an isolated current sheet, whose magnetic field profile is given in Equation
  (\ref{eq:single}), for three values of the ratio $k_{p0}/k_x$ of current sheet thickness to skin depth.
  Mode has a finite energy, with $A_{z1}$ and $\phi_1$ (or equivalently $\xi_x$ and $\xi_x^{\rm rec}$)
  both decaying exponentially in the exterior zone with uniform $B_{y0}$.   The growth rate drops toward
  the minimum $s_{\rm min}$ given by Equation (\ref{eq:smin}) as $k_{p0}/k_x$ approaches a maximum value $\simeq 1.86$.
  Red dotted curve shows $s_{\rm min}$ for $k_{p0} = 1.5k_x$; the growth rate (black curve) drops to this value
  at $k_y \simeq 0.54k_x$ and growth shuts off at larger $k_y$.    Modes with $s > 0$ also disappear when
  $k_y > k_{p0}/\bar\beta_0$;  this is plotted as the blue dotted line in the case $k_{p0} = 10^{-1/2}k_x$.\label{fig:single}}
\vskip 0.7in
\end{figure}

\begin{figure}
  \epsscale{0.55}
  \vskip -0.3in
  \plotone{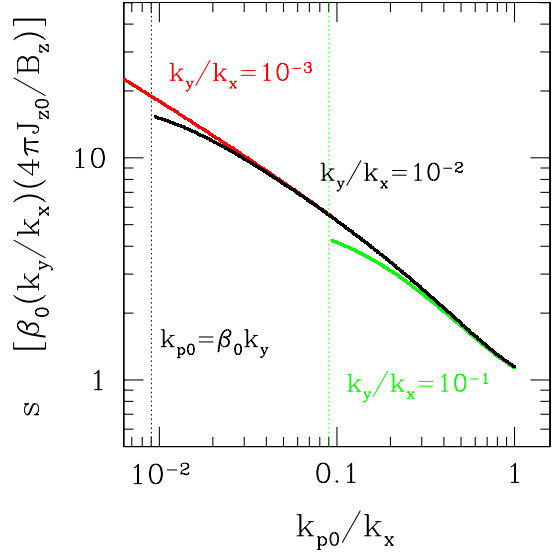}
\vskip -0.8in
\caption{Dependence of the growth rate of a tearing mode localized near a single current sheet on
  the ratio $k_{p0}/k_x$ of current sheet thickness to skin depth.    Curves correspond to the regime
  of small mode wavenumber $k_y$, with growth possible for $k_{p0} > \bar\beta_0k_y$.  Growth rate diverges
  for small current sheet thickness, approximately as $s \propto (k_x/k_{p0})^{1/2}$ for $k_{p0}/k_x < 0.1$.\label{fig:single2}}
\end{figure}

The isolated current sheet is taken to have the same current profile as previously (Equation (\ref{eq:back})),
but now restricted to a single reversal of $B_{y0}$, i.e.,
\be\label{eq:single}
B_{y0} =
\begin{cases}
  -{\cal B}_{\perp0} &  (x < -\pi/2k_x);\\
  {\cal B}_{\perp0}\sin(k_xx)\quad & (-\pi/2k_x < x < \pi/2k_x); \\
  +{\cal B}_{\perp0} & (x>\pi/2k_x).
\end{cases}
\ee

The perturbation equations (\ref{eq:Coulneut}) and (\ref{eq:Ampneut}) still apply in the current-free zones after taking
$k_x \rightarrow 0$.  Then the eigenfunctions for $\xi_x$, $\xi_x^{\rm rec}$ can be written in the form
$Ae^{ik_{\rm ex}x} + Be^{-ik_{\rm ex}x}$, where
\be\label{eq:kex}
k_{\rm ex}^2 = {(1-\widetilde s^2)(\bar\beta_0^{-2}+\widetilde s^2)\over (1+\widetilde s^2)^2}k_{p0}^2 - k_y^2
\ee
and $\widetilde s$ is defined after Equation (\ref{eq:approx}).  
The behavior may be oscillatory or exponential, but the mode energy converges only when the perturbation
is exponentially decaying at $|x| > \pi/2k_x$.  
This constraint implies a minimum growth rate, which is obtained by setting $k_{\rm ex} = 0$ in Equation (\ref{eq:kex}),
\be\label{eq:smin}
\widetilde s_{\rm min}^2 = {1 -2\varepsilon_k - \bar\beta_0^{-2} +
  \sqrt{(1 + \bar\beta_0^{-2})^2 + 8\varepsilon_k(\bar\beta_0^{-2}-1)}
  \over 2(1 + \varepsilon_k)},
\ee
where $\varepsilon_k \equiv k_y^2/k_{p0}^2$.   One finds real $s_{\rm min}$ only for $k_y > k_{p0}/\bar\beta_0$.

The growth rates of these finite-energy modes are plotted in Figure \ref{fig:single} for a few ratios of
current sheet thickness to skin depth.   One finds that as $k_{p0}/k_x$ approaches a limiting value close to 2
($\simeq 1.86$ for $k_y/k_x = 10^{-2}$), then $s \rightarrow s_{\rm min}$ and $k_{\rm ex} \rightarrow 0$.   These results are obtained, as before, by
iterating on $s$ and $d\phi_1/dx$ at $x = 0$, but now requiring that $d\ln A_{z1}/dx = d\ln\phi_1/dx =
-k_{\rm ex}\,{\rm sgn}(x)$ at $|x| = \pi/2k_x$.

The dependence of the growth rate on the thickness of the current sheet is shown in Figure \ref{fig:single2}.
One sees that $s$ diverges slowly, approximate as $s \propto k_x^{1/2}$, as the current sheet grows thinner
in comparison with the skin depth.

To summarize, 
we have demonstrated the existence of finite-energy and purely growing tearing perturbations, localized around an isolated
current sheet in a relativistic, quantizing magnetic field, that have similar growth rates to the lowest-order
modes found previously in the multiple-sheet configuration.  These growing modes localized around a single
current sheet exist only for a more restricted range of $k_{p0}/k_x$ and $k_y/k_{p0}$ than do the multiple-sheet modes.  The current sheet thickness $k_x^{-1}$ must be smaller than or comparable to
the skin depth, as must the reversal scale $k_y^{-1}$ of the excited mode along the tearing surface.
The increase in growth rate with decreasing gradient scale $k_x^{-1}$ supports the formation of strong, localized
current sheets during the non-linear development of the tearing mode, along the lines of the
\cite{syrovatskii71} model.

\begin{figure}
  \epsscale{0.55}
  \vskip -0.3in
  \plotone{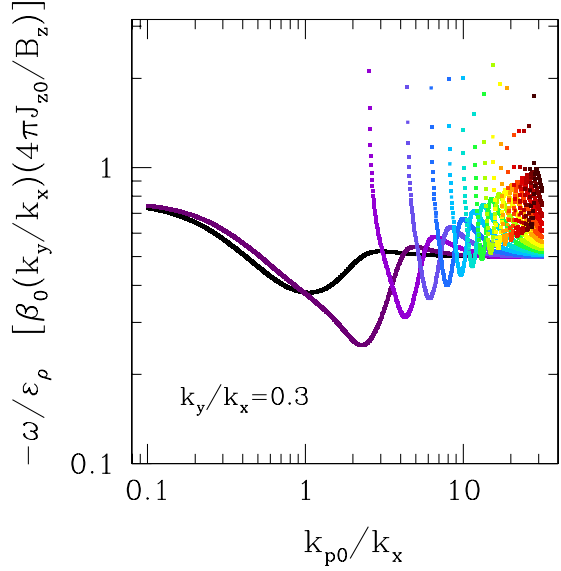}
\vskip -0.8in
\caption{Real frequency induced in the dispersion relation of Figure \ref{fig:dispersion} by a small charge imbalance
  in the plasma, as measured by the parameter $\varepsilon_\rho$ (Equation (\ref{eq:imbalance})).  Frequency is negative for
  $\varepsilon_\rho > 0$, and positive for $\varepsilon_\rho < 0$.  Growth rate $s = s_0$ is unperturbed
  to first order in $\varepsilon_\rho$.  Results are obtained by iterating on
  solutions to Equations (\ref{eq:CoulneutI}) and (\ref{eq:AmpneutI}) so as to maintain the boundary conditions (\ref{eq:bc2}).
  Colors mark the same tower of modes as in Figure \ref{fig:dispersion}, with $\xi_x^{\rm rec}$ node number $N_{\rm rec}$ increasing from 0
  (black) to 15 (dark red).\label{fig:frequency}}
\vskip 0.1in
\end{figure}

\section{Non-neutral Plasma}\label{s:nonneutral}

We now turn to the more general case of a non-neutral plasma, writing
\be\label{eq:imbalance}
n_0^\pm = \left(1 \pm \varepsilon_\rho\right){n_0\over 2}.
\ee
The parameter $\varepsilon_\rho$ is a measure of the charge imbalance, since
\be
\rho = q(n_0^+ - n_0^-) = \varepsilon_\rho qn_0.
\ee
The quantity appearing in Equation (\ref{eq:sprimeinv}) is now a complex number, meaning that we must take the frequency
to have a real component, $s \rightarrow s - i\omega$, and consider both $A_{z1}$ and $\phi_1$ to be complex numbers.  

The problem simplifies when we consider $\varepsilon_\rho$ to be a small parameter and treat the effect of the
charge imbalance as a perturbation to the charge-balanced solutions obtained in Section \ref{s:neutral}.
Then the perturbations to $A_{z1}$ and $\phi_1$ are purely imaginary,
\be
\delta A_{z1} = i\delta A_{z1}^I;  \quad\quad \delta\phi_1 = i\delta\phi_1^I,
\ee
and the real component of the growth rate perturbation vanishes to leading order, $s_0 \rightarrow s_0 - i\omega$.
To first order in $\omega$ and $\varepsilon_\rho$,
\be
   {n_0\over s'^2} =
   n_0\left\{{s_0^2-(\bar\beta_0ck_\parallel)^2\over [s_0^2 + (\bar\beta_0ck_\parallel)^2]^2} + if(s_0,k_\parallel)\right\},
\ee
where
\be
f(s_0,k_\parallel) = -2s_0^2{3(\bar\beta_0ck_\parallel)^2-s_0^2\over [s_0^2 + (\bar\beta_0ck_\parallel)^2]^3}\left({\omega\over s_0}\right)
- {2s_0\,\bar\beta_0ck_\parallel\over [s_0^2 + (\bar\beta_0ck_\parallel)^2]^2}\varepsilon_\rho
\ee
is proportional to the small parameters $\omega/s_0$, $\varepsilon_\rho$.  We will also need
\be
   {n_0(s_0-i\omega)^2\over s'^2} =
   n_0\left\{{s_0^2[s_0^2-(\bar\beta_0ck_\parallel)^2]\over [s_0^2 + (\bar\beta_0ck_\parallel)^2]^2} + ig(s_0,k_\parallel)\right\},
\ee
where
\be
g(s_0,k_\parallel) = -2s_0^2\left\{{[3s_0^2-(\bar\beta_0ck_\parallel)^2](\bar\beta_0ck_\parallel)^2\over
  [s_0^2 + (\bar\beta_0ck_\parallel)^2]^3}\left({\omega\over s_0}\right)
+ {s_0\,\bar\beta_0ck_\parallel\over [s_0^2 + (\bar\beta_0ck_\parallel)^2]^2}\varepsilon_\rho\right\}.
\ee

For the background current configuration, we return to the case of a sinusoidal
profile (Equation (\ref{eq:back})) and solve the mode Equations (\ref{eq:Coul}) and (\ref{eq:Amp})
expressed in terms of the displacement fields (\ref{eq:hydro}).
Writing $\xi_x \rightarrow \xi_x + i\delta\xi_x^I$ and
$\xi_x^{\rm rec} \rightarrow \xi_x^{\rm rec} + i\delta\xi_x^{{\rm rec},I}$, and subtracting out the
unperturbed eigenmode equations, we have
\ba\label{eq:CoulneutI}
         {d^2\,\delta\xi_x^I\over d(k_x x)^2} &=&
         {k_y^2\over k_x^2}\,\delta\xi_x^I -  (ck_\parallel)^2\left({k_{p0}^2\over k_x^2} + 1\right)
         \left\{{s_0^2-(\bar\beta_0ck_\parallel)^2\over [s_0^2 + (\bar\beta_0 ck_\parallel)^2]^2}\,\delta\xi_x^{{\rm rec},I} 
         + f(s_0,k_\parallel)\,\xi_x^{\rm rec}\right\},\nn
\ea
and
\ba\label{eq:AmpneutI}
         {d^2[B_{y0}(\delta\xi_x^I+\delta\xi_x^{{\rm rec},I})]\over d(k_xx)^2}
         &=& \left({k_y^2\over k_x^2} - 1\right)B_{y0}\,(\delta\xi_x^I + \delta\xi_x^{{\rm rec},I})  \nn &+&
   \left({k_{p0}^2\over k_x^2} + 1\right)\left\{{s_0^2[s_0^2-(\bar\beta_0ck_\parallel)^2]\over[s_0^2 + (\bar\beta_0ck_\parallel)^2]}
   B_{y0}\,\delta\xi_x^{{\rm rec},I} + g(s_0,k_\parallel)B_{y0}\,\xi_x^{\rm rec}\right\}.\nn
\ea
Here $\xi_x^{\rm rec}$ is obtained from the charge-neutral solution to Equations (\ref{eq:Coulneut}) and (\ref{eq:Ampneut}).
Only the non-ideal part of the electromagnetic perturbation is responsible for driving the oscillation, just as it played a key role
in driving the instability in the charge-neutral case.

\begin{figure}
  \epsscale{1.1}
  \vskip -0.3in
  \plottwo{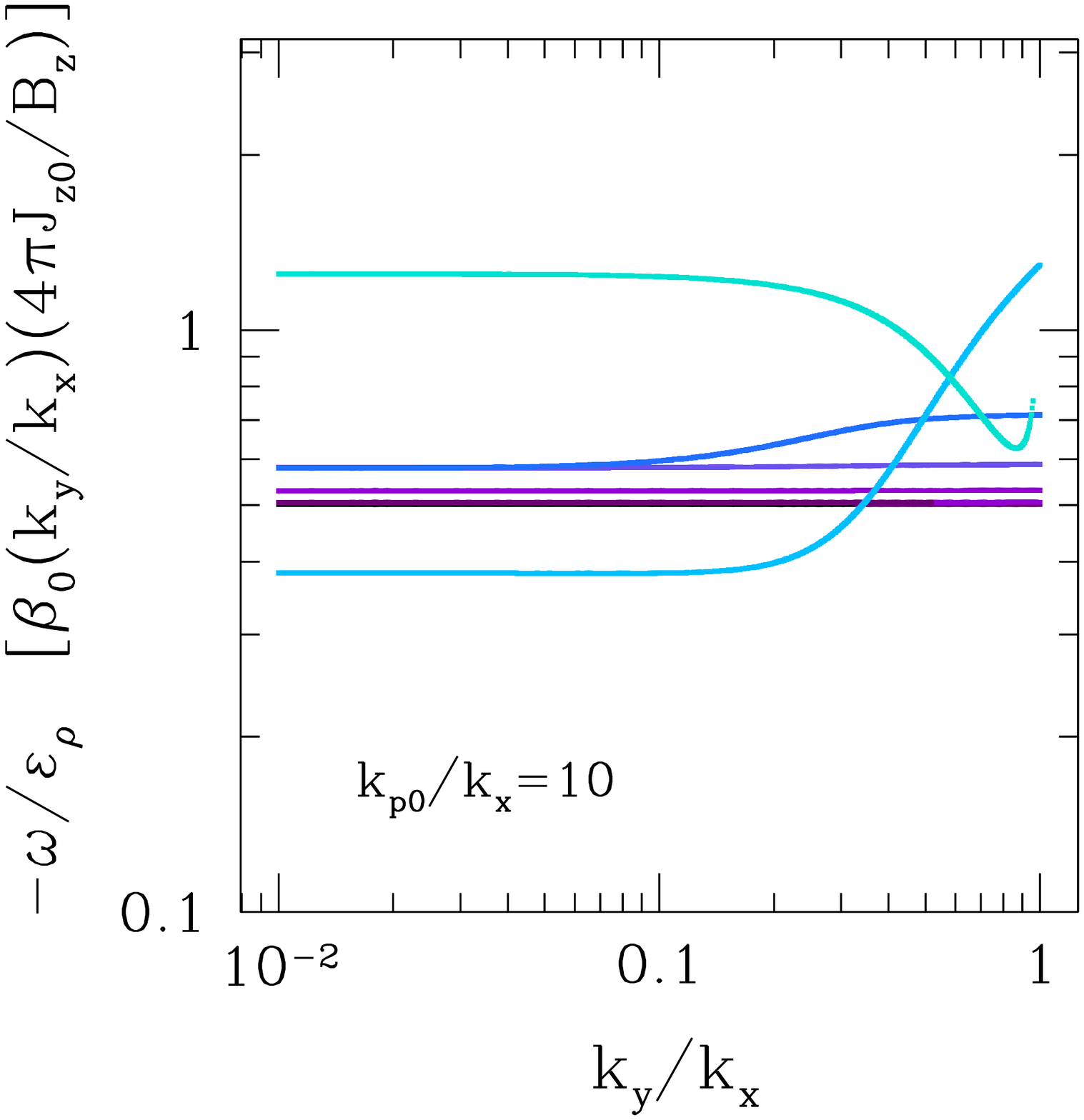}{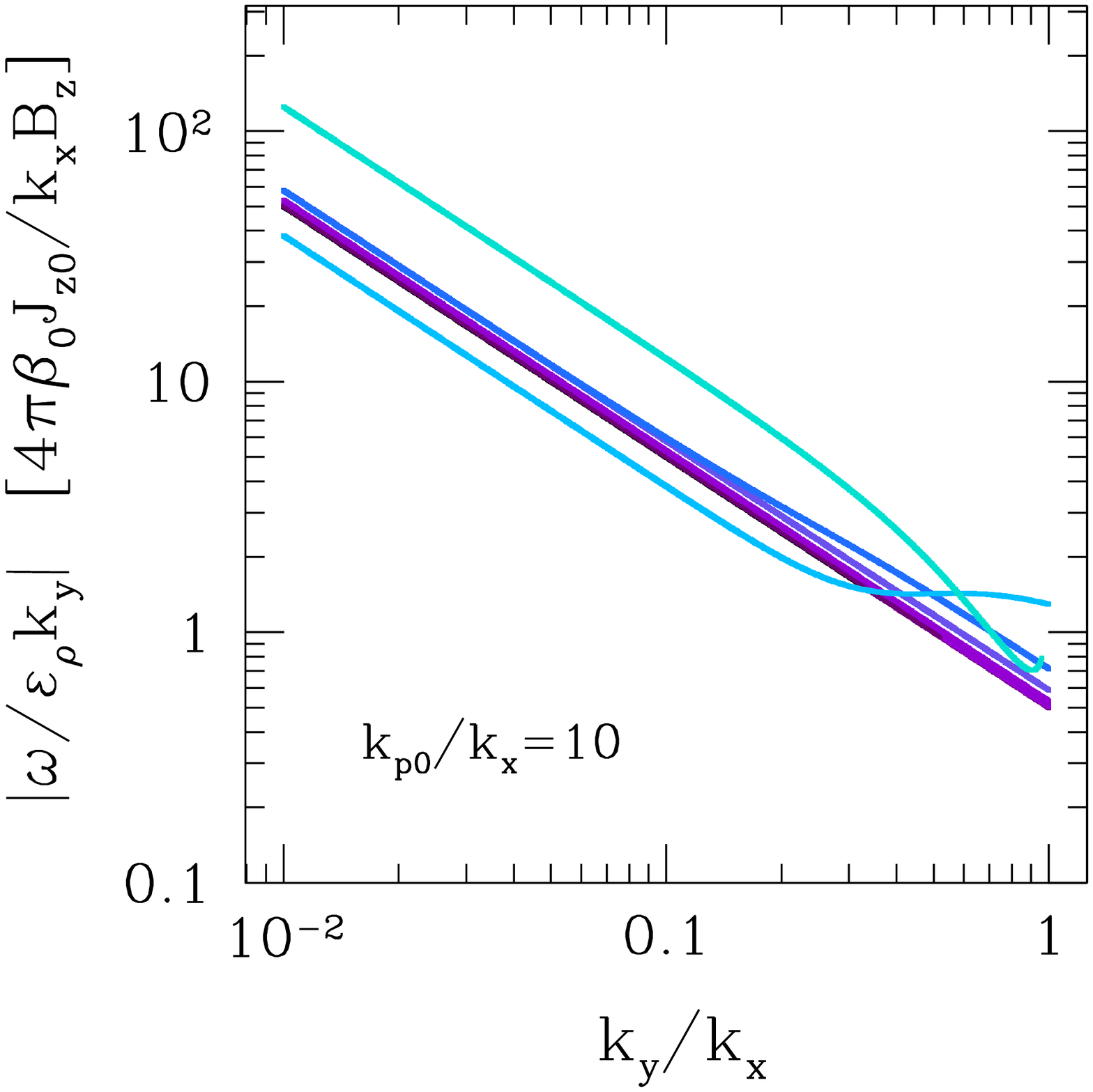}
\vskip -0.8in
\caption{Left panel:  Same as Figure \ref{fig:frequency}, but now $\omega$ is plotted versus the mode wavenumber $k_y$.
  Right panel:  Corresponding phase speed in the direction of the background non-potential magnetic field.\label{fig:frequency2}}
\vskip 0.1in
\end{figure}

We search for solutions to Equations (\ref{eq:CoulneutI}) and (\ref{eq:AmpneutI}) that have $\delta A_{z1}^I$ symmetric
about $x = 0$, $\delta\xi_x$ antisymmetric, and all perturbations vanishing at $x=0$:
\be\label{eq:bc2}
\delta A_{z1}^I(0) = {d\delta A_{z1}^I\over dx}(0) = \delta\xi_x(0) = 0.
\ee
The result is shown in Figures \ref{fig:frequency} and \ref{fig:frequency2}.

\section{Line Tying and Surface Plastic Flow}\label{s:linetying}

The tearing modes described in this paper involve a nearly incompressible and force-free rearrangement
of a strong guide magnetic field.  The transverse scale $\sim k_{p,\rm ex}^{-1}
\sim 0.1-100$ cm of the mode is tiny compared with the length $r \lesssim c/\Omega \sim 5\times 10^8\,
P_{-1}$ cm of the sheared magnetic field.  This field is embedded in the outer layers of a neutron star,
whose shear strength varies greatly with depth.  The transition from force-free to magnetoelastic equilibrium
is spread over a distance somewhat greater than the magnetospheric skin depth.

Line tying can reduce the growth rate of a tearing mode \citep{huang09}.  We find that its effect is limited
in the open circuit of a pulsar, given the extended length of the magnetic field lines.  More significant
is its effect on suppressing tearing modes in PIC simulations of pulsar magnetospheres,
especially simulations focusing on the polar cap region.  
To avoid such an artificial suppression, a numerical model must simultaneously cover the entire magnetosphere
as well as small-scale, tranverse structure in the open magnetic field bundle.

To understand the effect of line tying on reconnection in a force-free, relativistic plasma, consider
first the case of low  magnetization, $\sigma \ll 1$ (Alfv\'en speed $\ll c$).  
In the usual formulation of the problem, the sheared magnetic field  is tied at both ends, with
a length $\ell_\parallel$.  The growth rate is reduced when $k_z \sim \ell_\parallel^{-1}
\gtrsim k_\parallel \sim k_y {\cal B}_{\perp0}/B_{z0}$; then the mode becomes nearly force
free and $s \propto \ell_\parallel$ \citep{huang09}.

In the case of a quantizing magnetic field, as examined here, inertial effects are generally negligible and 
a growing tearing mode passes through a sequence of force-free equilibria.
The magnetic field in the open pulsar circuit is tied only at one end, its length
measured out to the light cylinder being comparable to $c/\Omega$.  The growth rate for infinite
$\ell_\parallel$ is $s \sim ck_\parallel \sim 2\Omega (k_y/k_x)$ (Equation (\ref{eq:slimit2})).
Line tying will therefore have an effect on the growth rate when $k_y \lesssim k_x/2$;  the reduction in
growth rate  $s \rightarrow s\cdot k_\parallel \ell_\parallel$ for $k_\parallel \ell_\parallel < 1$
then implies $s \rightarrow 4\Omega (k_y/k_x)^2$.   

High-order tearing of the slender open magnetic flux rope is insensitive to large-scale
curvature of the rope (Section \ref{s:quant}).  The growth rate is also insensitive to magnetic flaring:
when evaluated using a locally cartesian model, $s$ is proportional to $J_{z0}/B_{z0}$, which
is independent of radius
inside the speed-of-light cylinder.   Any curent
irregularities formed at small radius will propagate along the open field bundle and
be advected out into the pulsar wind.

The force-free structure of the tearing
mode is modified by solid stresses only in a narrow layer below the stellar surface.  The footpoints
of the magnetic field are not fixed in position;  instead the outer layers of the star
flow plastically with them \citep{li15,tyo17}, down to a critical depth that we now derive.
We start by balancing one component of the Maxwell stress with the yield stress in the solid,
\be
   {B_{y1}B_{z0}\over 4\pi} \sim \varepsilon_y\mu,
\ee
where $\varepsilon_y \sim 10^{-2}-10^{-1}$ is a temperature dependent yield strain
\citep{chugunov10} and the shear modulus is $\mu \simeq 0.12 (Ze)^2 (4\pi\rho/3Am_n)^{4/3}$ in a Coulomb solid 
composed of nuclei of charge $Ze$ and mass $Am_n$ \citep{strohmayer91}.  Then at the yielding depth,
the crustal mass density is
\be
\rho_y = 7.3\times 10^5\, {A\over Z^{3/2}}  {B_{z,15}^{9/8}\over \varepsilon_{y,-1}^{3/4}P_{-1}^{3/8}}
\left[{(k_p/k_x)f_J\over (1+2{\cal M}_{\pm})}\right]^{3/4}\quad {\rm g~cm^{-3}}.
\ee
Here, we have set the background
current to a fraction $f_J$ of $\rho_{\rm co}c$, that is, $k_xB_{y0} \sim 2f_J\Omega B_{z0}/c$,
and normalized $k_x$ to the plasma scale using $k_p^2 = 4\pi (1+2{\cal M}_\pm)|e\rho_{\rm co}|/m_ec^2$.
  
The depth of the yielding layer can be compared to the magnetospheric skin depth.  
Using the hydrostatic relation between depth and Fermi energy, $E_{{\rm F},e} = (A/Z)m_n g|z|$, where
$g$ is surface gravity, approximating
the degenerate electron gas as non-relativistic, and choosing a $^{26}$Fe composition,
one finds that the yielding depth is somewhat larger than the magnetospheric skin depth,
\be\label{eq:elastdepth}
k_p|z|_y = 2.8\,(1+2{\cal M}_\pm)^{1/4}{B_{z,12}^{5/4}\over \varepsilon_{y,-1}^{1/2}P_{-1}^{3/4}g_{14}}
\left[f_J{k_p\over k_x}\right]^{1/2}.
\ee
There is a moderate enhancement in this ratio when ${\cal M}_\pm \gg 1$.

\section{Summary and Application to Pulsars and Magnetars}\label{s:summary}

We have identified and characterized internal tearing modes in a quantizing and nearly force-free magnetic field.
The electric current profile contains both a smooth component and a high-wavenumber component that drives
internal tearing. The embedded plasma is collisionless and relativistic, experiencing negligible curvature drift.
Electron inertia plays a dominant role in the breakdown of magnetic flux freezing, although collisions
  are in practice expected to have non-negligible effects in magnetar magnetospheres.
Our results form the basis for a `top-down' approach
to pulsar radio emission, in which slow tearing operating away from the magnetic separatrix are
the ultimate driver of the emission -- as opposed to fast plasma oscillations that are excited in a
quasi-electrostatic gap \citep{levinson05,philippov20}.

Our main findings are as follows.  

1. A robust but slow instability is uncovered in the presence of multiple, interacting
tearing surfaces.  The growth rate is suppressed by the
inverse of the strong guide magnetic field running perpendicular to the plane of reconnection.
The dispersion relation is derived for a wide range of separations between tearing surfaces,
measured relative to the skin depth.  When this separation is large,
the peak growth rate is very nearly $s_0 \equiv 4\pi (k_y/k_x)J_{z0}/B_{z0}$, where $J_{z0}$ is
the high-wavenumber component of the
current density and $B_{z0}$ the guide field (Figure \ref{fig:dispersion} and Equation (\ref{eq:slimit})).
Faster growth is obtained when the background current is irregular on a scale smaller than the skin depth
$k_{p0}^{-1} \sim \langle\gamma_0^{-3}\rangle^{-1/2} (mc^2/4\pi n_0q^2)^{1/2}$.  
We find that current irregularities are susceptible to tearing over a wide range of scales.

2. Overstable modes are uncovered when the plasma carries net charge, as is expected
in the magnetosphere of a radio pulsar.
The real frequency $\omega$ is comparable to the growth rate $s$ times the relative imbalance $|n_0^+-n_0^-|/|n_0^++n_0^-|$
in the densities of positive and negative charges (Figure \ref{fig:frequency} and \ref{fig:frequency2}).

3. An isolated current sheet whose thickness is smaller than or comparable to the skin depth ($k_{p0}/k_x \lesssim 2$)  
is shown to sustain localized tearing modes (Figures \ref{fig:single} and \ref{fig:single2}).
Mode growth remains slow but scales with the ratio $k_{p0}/k_x$ of current sheet thickness
to skin depth as $s \sim (k_x/k_{p0})^{1/2}s_0$ for large $k_x/k_{p0}$.   This opens up a mechanism for runaway collapse
of the current sheet even in the presence of a strong guide field. 

4. Overtone tearing modes, showing one or more nodes in the non-ideal displacement field $\xi^{\rm rec}_x(x)$,
can be excited when neighboring tearing surfaces are separated by a distance exceeding
the skin depth, $k_x < k_{p0}$ (Figure \ref{fig:dispersion}).
The excitation of these overtone modes opens up a mechanism for creating small-scale
structure in the current surrounding a smooth extremum in the twist profile, as is needed to
redistribute twist across a magnetic flux bundle.

5. Line tying has a limited effect on the growth of small-scale tearing modes in the open circuit
of a rotating neutron star.  In significant part, that is because the growth length $c/s$ is comparable
to the length of the open-field bundle, but also because inertial effects have
a negligible effect on mode growth even in the absence of line tying.
We infer a smooth transition
from the force-free state of the magnetosphere to a weak magnetoelastic deformation of the upper
solid crust.  This transformation is concentrated at a depth somewhat larger than
the transverse size of the magnetospheric current irregularities (the magnetospheric skin depth;
Equation (\ref{eq:elastdepth})).

6. It nonetheless should be emphasized that
line tying will have the practical effect of suppressing tearing mode growth in a PIC simulation
of the open pulsar circuit that covers only a limited range $\Delta r$ of radius near the surface of the star:
growth rates will be suppressed by a factor $\sim \Omega\Delta r/c$ (see \citealt{huang09} for the case
of non-relativistic reconnection).

\subsection{Peculiarities of Tearing Modes in a Quantizing Magnetic Field}

1. The dynamics can be understood in terms of charge flow along separate magnetic flux tubes, which interact
via the Coulomb and Lorentz forces.   Charge density and current density perturbations are of comparable importance
in driving instability.  The vorticity field that is a central actor in non-relativistic reconnection is replaced
by the charge density field (Equations (\ref{eq:charcon}) and (\ref{eq:charcon2})).

2.  The modes uncovered here are a relativistic generalization of the double tearing mode, but with the distinction
that the non-ideal
and ideal displacements are comparable in magnitude throughout the bulk.  There is no separation
between a narrow non-ideal layer and a larger, hydromagnetic  deformation of the magnetic field.  Growth rates are obtained for
separations of the tearing surfaces up to $\sim 30$ times the skin depth.   In the perturbation equations,
the non-ideal displacement (see Equation (\ref{eq:hydro})) is seen to be the driver of mode growth
(Equations (\ref{eq:Coul}) and (\ref{eq:Amp})).   Similarly to
the double tearing mode \citep{bierwage05,wang07}, the displacement becomes very strongly peaked
in the bulk, as the surfaces $B_{y0} = 0$ become separated by more than a few skin depths
(Figure \ref{fig:mode}).

3. Also related to the non-ideal nature of the tearing instability is the failure of 
the `constant-$\Psi$' approximation,\footnote{In our notation, `constant-$A_{z1}$'.}
as applied to a single tearing surface in quantizing magnetic field.  In this approximation,
the energy of the perturbation diverges with distance from the current sheet.
This divergence is traced to the appearance of coordinated current and charge oscillations.  

4.  The tearing mode with fastest growth has the fewest number of nodes $N_{\rm rec}$ in the profile of
$\xi_x^{\rm rec}$ transverse to the tearing surface.  When $k_{p0} \gg k_x$ (the tearing surfaces are separated by
a distance much greater than the skin depth), a large tower of independent branches appears
in the dispersion relation, labeled by $N_{\rm rec} > 0$.  The
wavenumber of the $x$-oscillation lies in the range $k_x < k < k_{p0}$.  The
growth rates of these overtone modes are nearly as large as the growth rate of the fundamental mode
with zero nodes (Figure \ref{fig:dispersion}).

5. Nonlinear effects associated with magnetic island saturation \citep{rutherford73}
are expected to have a limited effect on mode growth,
given that the displacement is concentrated away from the tearing surface when $k_{p0} \gtrsim k_x$.

\subsection{Future Directions}

 The nonlinear development of tearing in a quantizing magnetic field is left to future work.
  Existing time-dependent calculations of tearing in a non-relativistic
  plasma show strong nonlinear growth when current sheets of opposing signs are in close proximity
  \citep{ishii02,bierwage05,zm11}.   This growth is driven by the expansion and eventual
  interaction of distinct populations of magnetic islands.
  Thus, the energy released by the instability described
  here may greatly exceed the simplest estimate extracted from linear theory.  Current irregularities
  forming near the skin depth also fit within a hierarchy of magnetic structures.  For example, the energy
  carried by a tearing mode feeding off a current irregularity of wavenumber $k_x$ will scale as
  $k_x^{-1}B_{y0}^2 \propto k_x^{-3}$,
  given that the seed current irregularity $J_{z0} \sim c k_x B_{y0}/4\pi$  is normalized
  by the mean current density $\sim \rho_{\rm co} c$ in the pulsar circuit.  The fast-growing
  overtone modes that are found for $k_x \ll k_{p0}$ give an indication of how longer-wavelength current
  irregularities may spawn smaller-scale magnetic structures.

Another avenue for further work involves generalizing the tearing dispersion relation derived here
(for a narrow top-hat particle distribution function, with
counterstreaming positive and negative charges) to particle distributions
with (i) a broader spread in momentum and (ii) a collective flow of positive and negative charges
(as expected in the parts of the pulsar polar cap experiencing a pair cascade).

 Heavy ions
  will also have an effect on the mode structure.  A space density
  $\sim \rho_{\rm co}/|Ze|$ is expected when the corotation charge is positive.
  In the absence of pairs,
  the characteristic scale of the tearing eigenmodes is increased by a factor $\sim (Am_p/m_e)^{1/2}$
  (here $Z$ and $A$ are the ion charge and mass in atomic units) but the growth rate and drift rates
  are otherwise independent of the mass of the particle carrying the corotation charge.   In the presence of pairs,
  the oscillatory behavior  of the overtone tearing eigenmodes will be modified compared with the case
  of negative corotation charge.

\subsection{Application to Radio Pulsars}

1. {\it Slow mode growth and pulsar flux variability.}
A role for the instability uncovered here in pulsar radiation emission is motivated by the presence of a strongly
inhomogeneous current profile in the open magnetic flux bundle of a rotating neutron star \citep{timokhin13,gralla17}.  In many cases, the
bundle shows non-axisymmetric features and an inhomogeneous distribution of magnetic twist that is known to
trigger fast resistive instabilities in tokamaks (e.g. the double tearing mode) and to induce a rapid relaxation
in the twist profile toward a local maximum \citep{white13}.  The vigor of such an instability is limited in the
pulsar context by the presence of a strong guide field:   
the peak growth rate is $\sim \Omega$,  the angular rotation
frequency of the star, which is much slower than the characteristic frequencies of particle and field oscillations
in a polar gap.  This is slow enough to be relevant to the dramatic variations that are observed in radio emission
over successive rotations \citep{graham-smith03}.

Establishing a connection between magnetospheric tearing and radio emission depends on
identifying a secondary instability that can induce charge clumping in current sheets.  Paper II
describes an overstability of propagating charge perturbations that are
localized near Debye-scale current sheets and are excited by a Cerenkov-like process.

2. {\it Sub-pulse drift.}
When the plasma carries net charge, as it must in the pulsar magnetosphere,
we uncover overstable tearing modes with a finite real frequency.
In a nearly axisymmetric configuration, as expected in a pulsar with small inclination between
the magnetic dipole and rotation axes, this overstability corresponds to a general drift
of the tearing modes in the azimuthal angle around the polar cap.  There is a promising
connection here with the secular `sub-pulse' drift that is observed in many pulsars \citep{deshpande01}.
Indeed, calculations of high-order (double, triple...) tearing modes in non-relativistic plasmas driven by
closely spaced tearing surfaces ($B_{y0} = 0$) show such an angular structure \citep{bierwage05}.

The magnitude of the drift rate depends on the particle density.  It is comparable to observed drift
rates in parts of the magnetosphere that have a low, or vanishing, $e^\pm$ pair density in the
absence of tearing (Paper II).

3. {\it Tearing at high and low current density.}  The slow tearing instability described here does not require
that the background current density $J_z$ exceed the limiting corotation charge flow $\rho_{\rm co}c$.
Self-consistent models of this particle flow indicate that
the charges begin to stream relativistically as $J_z$ approaches
$\rho_{\rm co}c$ \citep{timokhin13}.  The nonlinear development of slow tearing will increase
the current density locally and, at a given rotation rate,
expand the part of the polar cap experiencing $e^+$-$e^-$ pair cascades.  In the absence of
pair creation, the pulsar magnetosphere collapses from the configuration postulated by \cite{gj69}
to a static, charge-separated structure without relativistic particle flow \citep{krause85,chen14,philippov14}.
Small-scale tearing may therefore supplement general-relativistic
frame dragging \citep{mt92,philippov15} as a driver of charge starvation in the polar cap and
of activity as a pulsing radio source. 

\subsection{Application to Magnetars}

Neutron stars with magnetic fields strong enough to cause significant crustal yielding -- magnetars --
are frequently observed as intense sources of non-thermal X-ray, with a luminosity far exceeding what
can be supplied by the measured loss of rotational energy \citep{kb17}.  One infers the presence of
strong currents flowing through the closed magnetosphere and, from the simultaneous measurement of
thermal X-ray hotspots, a strong concentration of these currents on the magnetar surface.

The excitation of strong, localized magnetic shear during fault-like displacements of the crust has been
proposed as a trigger for runaway pair creation in magnetar X-ray flares \citep{td01,tyo17}.  A
path toward understanding how thermalization might occur in the stellar magnetosphere is provided
by the tearing instability described here, in combination with the secondary Cerenkov instability of shear Alfv\'en
waves investigated in Paper II.  High-frequency Alfv\'en waves that experience
rapid, in situ damping may be excited directly in the magnetosphere, as opposed to the indirect excitation
by crustal elastic modes \citep{blaes89}.  In the case of the giant magnetar X-ray flares, crustal elastic
modes are an inefficient source of magnetospheric dissipation (e.g. \citealt{td01,link14}).
Magnetar outbursts 
sometimes involve a brief non-thermal X-ray transient, followed by a slower relaxation lasting months
(e.g. \citealt{woods01,woods04}).
The slow component has been modeled in terms of ohmic decay of a localized magnetospheric current
\citep{beloborodov09}; alternatively, a global magneto-thermal-elastic model of the crust reveals
continued creep at crustal faults following an outburst \citep{tyo17},
from which one infers a role for internal tearing in powering delayed persistent
X-ray emission.  The instabilities described here and in Paper II may also play a role in the microphysical
development of a global reconnection event, involving the expulsion of a magnetic loop by the closed magnetosphere,
during a giant magnetar flare \citep{lyutikov06}.

\acknowledgements
An anonymous referee offered helpful comments on the presentation of our results.
We acknowledge the support of the Natural Sciences and Engineering Research Council of Canada (NSERC) through grant RGPIN-2017-06519.


\begin{thebibliography}{}
\bibitem[Basu \& Coppi(1981)]{basu81} Basu, B. \& Coppi, B.\ 1981, Physics of Fluids, 24, 465. doi:10.1063/1.863393
\bibitem[Beloborodov \& Thompson(2007)]{bt07} Beloborodov, A.~M. \& Thompson, C.\ 2007, \apj, 657, 967. doi:10.1086/508917
\bibitem[Beloborodov(2009)]{beloborodov09} Beloborodov, A.~M.\ 2009, \apj, 703, 1044. doi:10.1088/0004-637X/703/1/1044
%\bibitem[Bessho \& Bhattacharjee(2012)]{bessho12} Bessho, N. \& Bhattacharjee, A.\ 2012, \apj, 750, 129. doi:10.1088/0004-637X/750/2/129
\bibitem[Bierwage et al.(2005)]{bierwage05} Bierwage, A., Hamaguchi, S., Wakatani, M., et al.\ 2005, \prl, 94, 065001. doi:10.1103/PhysRevLett.94.065001
\bibitem[Biskamp(2005)]{biskamp05} Biskamp, D.\ 2005, Magnetic Reconnection in Plasmas, by Dieter Biskamp, Cambridge, UK: Cambridge University Press, 2005
\bibitem[Blaes et al.(1989)]{blaes89} Blaes, O., Blandford, R., Goldreich, P., et al.\ 1989, \apj, 343, 839. doi:10.1086/167754
\bibitem[Boozer(2020)]{boozer20} Boozer, A.~H.\ 2020, Physics of Plasmas, 27, 102305. doi:10.1063/5.0014107
\bibitem[Browning et al.(1986)]{browning86} Browning, P.~K., Sakurai, T., \& Priest, E.~R.\ 1986, \aap, 158, 217
\bibitem[Bucciantini et al.(2011)]{bucciantini11} Bucciantini, N., Arons, J., \& Amato, E.\ 2011, \mnras, 410, 381. doi:10.1111/j.1365-2966.2010.17449.x
\bibitem[Cerutti et al.(2013)]{cerutti13} Cerutti, B., Werner, G.~R., Uzdensky, D.~A., et al.\ 2013, \apj, 770, 147. doi:10.1088/0004-637X/770/2/147
\bibitem[Cerutti \& Philippov(2017)]{cerutti17} Cerutti, B. \& Philippov, A.~A.\ 2017, \aap, 607, A134. doi:10.1051/0004-6361/201731680
\bibitem[Chen \& Beloborodov(2014)]{chen14} Chen, A.~Y. \& Beloborodov, A.~M.\ 2014, \apjl, 795, L22. doi:10.1088/2041-8205/795/1/L22
\bibitem[Chen \& Beloborodov(2017)]{chen17} Chen, A.~Y. \& Beloborodov, A.~M.\ 2017, \apj, 844, 133. doi:10.3847/1538-4357/aa7a57
%\bibitem[Coppi et al.(1976)]{coppi76} Coppi, B., Galvao, R., Pellat, R., et al.\ 1976, Soviet Journal of Plasma Physics, 2, 533
\bibitem[Chugunov \& Horowitz(2010)]{chugunov10} Chugunov, A.~I. \& Horowitz, C.~J.\ 2010, \mnras, 407, L54. doi:10.1111/j.1745-3933.2010.00903.x
\bibitem[de Jager(2007)]{dejager07} de Jager, O.~C.\ 2007, \apj, 658, 1177. doi:10.1086/511950
\bibitem[Deshpande \& Rankin(2001)]{deshpande01} Deshpande, A.~A. \& Rankin, J.~M.\ 2001, \mnras, 322, 438. doi:10.1046/j.1365-8711.2001.04079.x
\bibitem[Drake \& Lee(1977)]{drake_lee77} Drake, J.~F. \& Lee, Y.~C.\ 1977, Physics of Fluids, 20, 1341. doi:10.1063/1.862017
\bibitem[Galeev \& Zeleny{\v{i}}(1976)]{galeev76} Galeev, A.~A. \& Zeleny{\v{i}}, L.~M.\ 1976, Soviet Journal of Experimental and Theoretical Physics, 43, 1113
\bibitem[Goldreich \& Julian(1969)]{gj69} Goldreich, P. \& Julian, W.~H.\ 1969, \apj, 157, 869. doi:10.1086/150119
\bibitem[Gourgouliatos et al.(2016)]{gourg16} Gourgouliatos, K.~N., Wood, T.~S., \& Hollerbach, R.\ 2016, Proceedings of the National Academy of Science, 113, 3944
\bibitem[Graham-Smith(2003)]{graham-smith03} Graham-Smith, F.\ 2003, Reports on Progress in Physics, 66, 173. doi:10.1088/0034-4885/66/2/203
\bibitem[Gralla et al.(2017)]{gralla17} Gralla, S.~E., Lupsasca, A., \& Philippov, A.\ 2017, \apj, 851, 137. doi:10.3847/1538-4357/aa978d
\bibitem[Gruzinov(2007)]{gruzinov07} Gruzinov, A.\ 2007, \apjl, 667, L69. doi:10.1086/519839
\bibitem[Hibschman \& Arons(2001)]{hibschman01} Hibschman, J.~A. \& Arons, J.\ 2001, \apj, 554, 624. doi:10.1086/321378
%\bibitem[Heyvaerts \& Priest(1984)]{hp84} Heyvaerts, J. \& Priest, E.~R.\ 1984, \aap, 137, 63
\bibitem[Hoshino(2021)]{hoshino21} Hoshino, M.\ 2021, Physics of Plasmas, 28, 062106. doi:10.1063/5.0050389
\bibitem[Huang \& Zweibel(2009)]{huang09} Huang, Y.-M. \& Zweibel, E.~G.\ 2009, Physics of Plasmas, 16, 042102. doi:10.1063/1.3103789
\bibitem[Ishii et al.(2002)]{ishii02} Ishii, Y., Azumi, M., \& Kishimoto, Y.\ 2002, \prl, 89, 205002. doi:10.1103/PhysRevLett.89.205002
%\bibitem[Huang et al.(2010)]{huang10} Huang, Y.-M., Bhattacharjee, A., \& Zweibel, E.~G.\ 2010, Physics of Plasmas, 17, 055707. doi:10.1063/1.3398486
\bibitem[Kaspi \& Beloborodov(2017)]{kb17} Kaspi, V.~M. \& Beloborodov, A.~M.\ 2017, \araa, 55, 261
%\bibitem[Komissarov et al.(2007)]{komissarov07} Komissarov, S.~S., Barkov, M., \& Lyutikov, M.\ 2007, \mnras, 374, 415. doi:10.1111/j.1365-2966.2006.11152.x
\bibitem[Krause-Polstorff \& Michel(1985)]{krause85} Krause-Polstorff, J. \& Michel, F.~C.\ 1985, \mnras, 213, 43. doi:10.1093/mnras/213.
\bibitem[Levinson et al.(2005)]{levinson05} Levinson, A., Melrose, D., Judge, A., et al.\ 2005, \apj, 631, 456. doi:10.1086/432498
\bibitem[Li \& Beloborodov(2015)]{li15} Li, X. \& Beloborodov, A.~M.\ 2015, \apj, 815, 25. doi:10.1088/0004-637X/815/1/25
\bibitem[Link(2014)]{link14} Link, B.\ 2014, \mnras, 441, 2676. doi:10.1093/mnras/stu584
\bibitem[Lyutikov(2003)]{lyutikov03} Lyutikov, M.\ 2003, \mnras, 346, 540. doi:10.1046/j.1365-2966.2003.07110.x
\bibitem[Lyutikov(2006)]{lyutikov06} Lyutikov, M.\ 2006, \mnras, 367, 1594. doi:10.1111/j.1365-2966.2006.10069.x
\bibitem[Muslimov \& Tsygan(1992)]{mt92} Muslimov, A.~G. \& Tsygan, A.~I.\ 1992, \mnras, 255, 61. doi:10.1093/mnras/255.1.61
\bibitem[Nalewajko et al.(2016)]{nalewajko16} Nalewajko, K., Zrake, J., Yuan, Y., et al.\ 2016, \apj, 826, 115. doi:10.3847/0004-637X/826/2/115
\bibitem[Ottaviani \& Porcelli(1993)]{ottaviani93} Ottaviani, M. \& Porcelli, F.\ 1993, \prl, 71, 3802. doi:10.1103/PhysRevLett.71.3802
\bibitem[Parker(1972)]{parker72} Parker, E.~N.\ 1972, \apj, 174, 499. doi:10.1086/151512
\bibitem[Philippov \& Spitkovsky(2014)]{philippov14} Philippov, A.~A. \& Spitkovsky, A.\ 2014, \apjl, 785, L33. doi:10.1088/2041-8205/785/2/L33
\bibitem[Philippov et al.(2015)]{philippov15} Philippov, A.~A., Cerutti, B., Tchekhovskoy, A., et al.\ 2015, \apjl, 815, L19. doi:10.1088/2041-8205/815/2/L19
\bibitem[Philippov et al.(2019)]{philippov19} Philippov, A., Uzdensky, D.~A., Spitkovsky, A., et al.\ 2019, \apjl, 876, L6. doi:10.3847/2041-8213/ab1590
\bibitem[Philippov et al.(2020)]{philippov20} Philippov, A., Timokhin, A., \& Spitkovsky, A.\ 2020, \prl, 124, 245101. doi:10.1103/PhysRevLett.124.245101
\bibitem[Priest \& Forbes(2000)]{priest00} Priest, E. \& Forbes, T.\ 2000, Magnetic Reconnection, by Eric Priest and Terry Forbes, pp. 612. ISBN 0521481791. Cambridge, UK: Cambridge University Press, June 2000., 612
\bibitem[Ruderman \& Sutherland(1975)]{rs75} Ruderman, M.~A. \& Sutherland, P.~G.\ 1975, \apj, 196, 51. doi:10.1086/153393
\bibitem[Russell et al.(2015)]{russell15} Russell, A.~J.~B., Yeates, A.~R., Hornig, G., et al.\ 2015, Physics of Plasmas, 22, 032106. doi:10.1063/1.4913489
\bibitem[Rutherford(1973)]{rutherford73} Rutherford, P.~H.\ 1973, Physics of Fluids, 16, 1903. doi:10.1063/1.1694232
\bibitem[Sironi \& Spitkovsky(2011)]{sironi11} Sironi, L. \& Spitkovsky, A.\ 2011, \apj, 741, 39. doi:10.1088/0004-637X/741/1/39
\bibitem[Sironi \& Spitkovsky(2014)]{sironi14} Sironi, L. \& Spitkovsky, A.\ 2014, \apjl, 783, L21. doi:10.1088/2041-8205/783/1/L21
\bibitem[Spitkovsky(2006)]{spitkovsky06} Spitkovsky, A.\ 2006, \apjl, 648, L51. doi:10.1086/507518
\bibitem[Strohmayer et al.(1991)]{strohmayer91} Strohmayer, T., Ogata, S., Iyetomi, H., et al.\ 1991, \apj, 375, 679. doi:10.1086/170231
\bibitem[Syrovatski{\v{i}}(1971)]{syrovatskii71} Syrovatski{\v{i}}, S.~I.\ 1971, Soviet Journal of Experimental and Theoretical Physics, 33, 933
\bibitem[Taylor(1986)]{taylor86} Taylor, J.~B.\ 1986, Reviews of Modern Physics, 58, 741. doi:10.1103/RevModPhys.58.741
\bibitem[Thompson \& Duncan(2001)]{td01} Thompson, C. \& Duncan, R.~C.\ 2001, \apj, 561, 980. doi:10.1086/323256
\bibitem[Thompson et al.(2002)]{tlk02} Thompson, C., Lyutikov, M., \& Kulkarni, S.~R.\ 2002, \apj, 574, 332. doi:10.1086/340586
\bibitem[Thompson(2008)]{thompson08} Thompson, C.\ 2008, \apj, 688, 1258. doi:10.1086/592263
  \bibitem[Thompson et al.(2017)]{tyo17} Thompson, C., Yang, H., \& Ortiz, N.\ 2017, \apj, 841, 54. doi:10.3847/1538-4357/aa6c30
  \bibitem[Thompson \& Kostenko(2020)]{tk20} Thompson, C. \& Kostenko, A.\ 2020, \apj, 904, 184. doi:10.3847/1538-4357/abbe87
  \bibitem[Thompson(2021)]{t21} Thompson, C. 2021, arXiv preprint
  \bibitem[Timokhin \& Arons(2013)]{timokhin13} Timokhin, A.~N. \& Arons, J.\ 2013, \mnras, 429, 20. doi:10.1093/mnras/sts298
\bibitem[Timokhin \& Harding(2019)]{th19} Timokhin, A.~N. \& Harding, A.~K.\ 2019, \apj, 871, 12. doi:10.3847/1538-4357/aaf050
\bibitem[Wang et al.(2007)]{wang07} Wang, Z.~X., Wang, X.~G., Dong, J.~Q., et al.\ 2007, \prl, 99, 185004. doi:10.1103/PhysRevLett.99.185004
\bibitem[White(2013)]{white13} White, R.~B.\ 2013, The Theory of Toroidally Confined Plasmas (Third Edition). (World Scientific: Singapore) 916 pp
\bibitem[Woods et al.(2001)]{woods01} Woods, P.~M., Kouveliotou, C., G{\"o}{\v{g}}{\"u}{\c{s}}, E., et al.\ 2001, \apj, 552, 748. doi:10.1086/320571
\bibitem[Woods et al.(2004)]{woods04} Woods, P.~M., Kaspi, V.~M., Thompson, C., et al.\ 2004, \apj, 605, 378. doi:10.1086/382233
\bibitem[Zelenyi \& Krasnoselskikh(1979)]{zelenyi79} Zelenyi, L.~M. \& Krasnoselskikh, V.~V.\ 1979, \sovast, 23, 460
\bibitem[Zenitani \& Hoshino(2008)]{zenitani08} Zenitani, S. \& Hoshino, M.\ 2008, \apj, 677, 530. doi:10.1086/528708
\bibitem[Zhang \& Ma(2011)]{zm11} Zhang, C.~L. \& Ma, Z.~W.\ 2011, Physics of Plasmas, 18, 052303. doi:10.1063/1.3581064
  
\end{thebibliography}
\end{document}